\documentclass[prd,nofootinbib,preprint,superscriptaddress]{revtex4}
\pdfoutput=1
\usepackage[T1]{fontenc}
\usepackage{amsmath,amssymb}
\usepackage{epsfig}
\usepackage{float}
\usepackage{graphicx}
\usepackage[usenames,dvipsnames]{color}
\usepackage{subfigure}
\usepackage{comment}
\usepackage{multirow}
\usepackage{slashed}
\usepackage[colorlinks,citecolor=blue]{hyperref}
\usepackage{pdfpages}
\usepackage{color}
\newcommand{\Tr}[1]{\text{Tr}\big[#1\big]}

\def\be{\begin{equation}}
\def\ee{\end{equation}}
\def\bsp#1\esp{\begin{split}#1\end{split}}

\def\bpm{\begin{pmatrix}}
	\def\epm{\end{pmatrix}}

\begin{document}
\title{Observing Left-Right Symmetry in the \\ Cosmic Microwave Background }

\author{Debasish Borah}
\email{dborah@iitg.ac.in}
\affiliation{Department of Physics, Indian Institute of Technology Guwahati, Assam 781039, India}

\author{Arnab Dasgupta}
\email{arnabdasgupta@protonmail.ch}
\affiliation{Institute of Convergence Fundamental Studies , Seoul-Tech, Seoul 139-743, Korea}

\author{Chayan Majumdar}
\email{chayan@phy.iitb.ac.in}
\affiliation{Department of Physics, Indian Institute of Technology Bombay, Mumbai 400076, India}

\author{Dibyendu Nanda}
\email{dibyendu.nanda@iitg.ac.in}
\affiliation{Department of Physics, Indian Institute of Technology Guwahati, Assam 781039, India}

\begin{abstract}
We consider the possibility of probing left-right symmetric model (LRSM) via cosmic microwave background (CMB). We adopt the minimal LRSM with Higgs doublets, also known as the doublet left-right model (DLRM), where all fermions including the neutrinos acquire masses only via their couplings to the Higgs bidoublet. Due to the Dirac nature of light neutrinos, there exist additional relativistic degrees of freedom which can thermalise in the early universe by virtue of their gauge interactions corresponding to the right sector. We constrain the model from Planck 2018 bound on the effective relativistic degrees of freedom and also estimate the prospects for planned CMB Stage IV experiments to constrain the model further. We find that $W_R$ boson mass below 4.06 TeV can be ruled out from Planck 2018 bound at $2\sigma$ CL in the exact left-right symmetric limit which is equally competitive as the LHC bounds from dijet resonance searches. On the other hand Planck 2018 bound at $1\sigma$ CL can rule out a much larger parameter space out of reach of present direct search experiments, even in the presence of additional relativistic degrees of freedom around the TeV corner. We also study the consequence of these constraints on dark matter in DLRM by considering a right handed real fermion quintuplet to be the dominant dark matter component in the universe.
\end{abstract}

\maketitle

\section{Introduction}
Left-right symmetric models (LRSM) \cite{Pati:1974yy, Mohapatra:1974hk, Mohapatra:1974gc, Senjanovic:1975rk, Senjanovic:1978ev, Mohapatra:1977mj, Mohapatra:1980qe, Mohapatra:1980yp, Lim:1981kv, Gunion:1989in,Deshpande:1990ip, FileviezPerez:2008sr} have been one of the most popular beyond standard model (BSM) frameworks studied in the literature. Here the gauge symmetry of the SM is extended to $ \rm SU(3)_c\times SU(2)_L\times SU(2)_R\times U(1)_{B-L}$ so that the right-handed fermions (which are singlet in SM) can form doublets under the new $SU(2)_R$. This not only makes the inclusion of right-handed neutrino automatic, but also puts the left and right-handed fermions on equal footing. Incorporating an additional discrete symmetry (or left-right parity) ensures that the theory is invariant under $SU(2)_L \leftrightarrow SU(2)_R$. Thus, the model not only explains the origin of parity violation in electroweak interactions through spontaneous breaking of a parity symmetric theory at high energy scale but also incorporates right handed neutrino, crucial to generate light neutrino masses and mixing observed at neutrino oscillation experiments \cite{Mohapatra:2005wg, Tanabashi:2018oca}. Apart from the possibility of realising left-right symmetry as an intermediate symmetry in popular grand unified theories (GUT) like SO(10), a TeV scale realisation of LRSM can also have very interesting consequences at collider experiments like the large hadron collider (LHC) \cite{Aaboud:2017efa,Aaboud:2017yvp,Sirunyan:2016iap,Khachatryan:2016jww,Sirunyan:2018mpc, Aad:2019hjw, Sirunyan:2019vgj}.

Conventionally, the very first proposals and studies of LRSM  \cite{Pati:1974yy, Mohapatra:1974hk, Mohapatra:1974gc, Senjanovic:1975rk, Senjanovic:1978ev} considered a scalar bidoublet for generating fermion masses and also for electroweak symmetry breaking whereas a pair of scalar doublets were introduced for the purpose of left-right symmetry breaking at high energy scale. A very recent detailed study of this model can be found in \cite{Bernard:2020cyi}. On the other hand, the LRSM proposals put forward later \cite{Mohapatra:1980qe, Mohapatra:1980yp, Lim:1981kv, Gunion:1989in,Deshpande:1990ip, FileviezPerez:2008sr} received much more attention due to the possibility of seesaw origin of light neutrino masses through a combination of type I seesaw \cite{Minkowski:1977sc, GellMann:1980vs, Mohapatra:1979ia, Schechter:1980gr} and type II seesaw \cite{Mohapatra:1980yp, Lazarides:1980nt, Wetterich:1981bx, Schechter:1981cv, Brahmachari:1997cq} or type III seesaw \cite{Foot:1988aq}. In the doublet left-right model (DLRM), in its minimal version, there is no such seesaw mechanism as all fermions including neutrinos acquire Dirac masses by virtue of their couplings to the bidoublet scalar. While generating sub-eV neutrino mass in this fashion requires relevant Yukawa couplings at the level of $ < 10^{-12}$, we adopt this minimal scenario to study some of the interesting phenomenological consequences. Radiative generation of light Dirac neutrinos in different left-right symmetric models have also been discussed over last few decades \cite{Babu:1988yq, Ma:1989tz, Ma:2016mwh, Borah:2016lrl, Borah:2016hqn, Borah:2017leo, Ma:2017kgb, Chavez:2019yal} which may provide a UV completion of the minimal DLRM we discuss here. Since such UV completions do not drastically change the conclusions we reach in the present work, we stick to the DLRM for the sake of simplicity.

The Dirac nature of light neutrinos in DLRM gives rise to additional relativistic degrees of freedom which can be thermalised in the early universe due to their gauge interactions mediated by right sector gauge bosons. Such additional light degrees of freedom can be probed by precise measurements of the cosmic microwave background (CMB) anisotropies. Recent 2018 data from the CMB measurement by the Planck satellite \cite{Aghanim:2018eyx} suggests that the effective degrees of freedom for neutrinos as 
\begin{eqnarray}
{\rm
N_{eff}= 2.99^{+0.34}_{-0.33}
}
\label{Neff}
\end{eqnarray}
at $2\sigma$ or $95\%$ CL including baryon acoustic oscillation (BAO) data. At $1\sigma$ CL it becomes more stringent to $N_{\rm eff} = 2.99 \pm 0.17$. Both these bounds are consistent with the standard model (SM) prediction $N^{\rm SM}_{\rm eff}=3.045$ \cite{Mangano:2005cc, Grohs:2015tfy, deSalas:2016ztq}. Upcoming CMB Stage IV (CMB-S4) experiments are expected to put much more stringent bounds than Planck due to their potential of probing all the way down to $\Delta N_{\rm eff}=N_{\rm eff}-N^{\rm SM}_{\rm eff} = 0.06$ \cite{Abazajian:2019eic}. We use the existing constraints and put strong limits on the scale of left-right symmetry or equivalently the right sector gauge bosons $W_R, Z_R$. For comparison, we also check the corresponding bounds for left-right asymmetric scenario by considering different $SU(2)_R$ gauge couplings. Interestingly, we find that the bounds on $W_R, Z_R$ mass from Planck 2018 bound on $\Delta N_{\rm eff}$ at $2\sigma$ CL are equally competitive as the latest LHC bounds \cite{Aaboud:2017yvp,Sirunyan:2016iap,Aad:2019hjw} and much stronger that the corresponding bounds from flavour physics \cite{Zhang:2007da}. On the other hand the same Planck 2018 bound at $1\sigma$ CL can rule out a much larger mass window for $W_R, Z_R$ out of reach of present collider experiments. In fact CMB-S4 will be able to probe a much larger region of $W_R, Z_R$ masses out of existing collider's reach and hence can probe or rule out the minimal model. Since there have been a few recent studies on gauged $B-L$ model with light Dirac neutrinos \cite{Abazajian:2019oqj, FileviezPerez:2019cyn, Nanda:2019nqy, Han:2020oet}and corresponding constraints due to Planck 2018 bound on $\Delta N_{\rm eff}$, we also reproduce the corresponding parameter space in gauged $B-L$ model and compare with the one obtained in DLRM. We point out the important difference due to the restricted range of DLRM gauge couplings $g_R, g_{BL}$ unlike that in gauged $B-L$ model. We also show the impact of these constraints on dark matter (DM) parameter space in DLRM by considering a right handed fermion quintuplet to be the dominant component of DM which can thermalise by virtue of its interactions with SM mediated by right sector gauge bosons. We calculate the parameter space allowed from observed DM relic and find the leftover parameter space after applying the $\Delta N_{\rm eff}$ bound. Finally, we comment on the more stringent Planck 2018 $1\sigma$  bound which can be satisfied if more light fields below the scale of left-right symmetry breaking in addition to SM plus three right handed neutrinos exist. In fact, we show that DLRM with right handed fermion quintuplet DM can give rise to sufficient number of additional degrees of freedom to serve this purpose. 

This paper is organised as follows. In section \ref{sec1}, we discuss the doublet left-right symmetric model followed by discussion of additional relativistic degrees of freedom due to light Dirac neutrinos in section \ref{sec2}. In section \ref{sec2a} we briefly discuss dark matter in DLRM particularly focusing on fermion quintuplet DM followed by results and discussion in section \ref{sec3}. We finally conclude in section \ref{sec4}.

\section{The DLRM}
\label{sec1}
\begin{table}
\begin{center}
\begin{tabular}{|c|c|}
\hline
Particles & $SU(3)_c \times SU(2)_L \times SU(2)_R \times U(1)_{B-L}$   \\
\hline
$Q_L=\begin{pmatrix}u_{L}\\
d_{L}\end{pmatrix}$ & $(3, 2, 1, \frac{1}{3})$  \\
$Q_R=\begin{pmatrix}u_{R}\\
d_{R}\end{pmatrix}$ & $(3, 1,2, \frac{1}{3})$   \\
$\ell_L=\begin{pmatrix}\nu_{L}\\
e_{L}\end{pmatrix}$ & $(1, 2, 1, -1)$  \\
$\ell_R=\begin{pmatrix}\nu_{R}\\
e_{R}\end{pmatrix}$ & $(1, 1, 2, -1)$ \\
\hline
\end{tabular}
\end{center}
\caption{Fermionic fields of the present Model including
the SM fermions.}
\label{tab:1}
\end{table}
\begin{table}
\begin{center}
\begin{tabular}{|c|c|}
\hline
Particles & $SU(3)_c \times SU(2)_L \times SU(2)_R \times U(1)_{B-L}$   \\
\hline
$\Phi=\begin{pmatrix}\phi^0 \,\,\,\,\,\, \phi^{\prime\, +} \\
\phi^- \,\,\,\,\,\, \phi^{\prime\, 0}\end{pmatrix}$ & $(1,2,2,0)$  \\
\hline
$\chi_L$ & $(1, 2, 1, 1)$ \\
$\chi_R$ & $(1, 1, 2, 1)$ \\
\hline
\end{tabular}
\end{center}
\caption{Scalar fields and their corresponding charges under all
the symmetry groups.}
\label{tab:2}
\end{table}

We briefly discuss the doublet left-right symmetric model in this section. The fermion and scalar content of the model are given in table \ref{tab:1} and \ref{tab:2} respectively. The relevant Yukawa Lagrangian giving masses to the three generations of leptons is given by,
	\begin{equation}\label{eq4}
	\mathcal{L}=h_{ij}\overline{l}_{L,i}\Phi l_{R,j}+\widetilde{h_{ij}}\overline{l}_{L,i}\widetilde{\Phi}l_{R,j}+{\rm h.c},
	\end{equation}
	where the indices $i,j=1,2,3$ represent the family indices for the three generations of fermions, $\widetilde{\Phi}=\tau_2\phi^*\tau_2$ and $\tau_2$ is Pauli matrix. The gauge structure of the model prevents any renormalisable Yukawa couplings involving the scalar doublets $\chi_{L, R}$. The scalar potential $V_{\text{scalar}}$ is given by \cite{Bernard:2020cyi}
{\small \be\bsp
		V_{\text{scalar}} & = -\mu_1^2 \Tr{\Phi^{\dagger} \Phi}- \mu_2^2\Tr{\Phi^\dagger\tilde{\Phi} + \tilde{\Phi}^\dagger\Phi}- \mu_3^2 (\chi_L^\dagger\chi_L + \chi_R^\dagger \chi_R)+ \lambda_1\Big(\Tr{\Phi^\dagger\Phi}\Big)^2 \\ 
		&\ + \lambda_2\Big\{ \Big(\Tr{\Phi^\dagger\tilde{\Phi}}\Big)^2+ \Big(\Tr{\tilde{\Phi}^\dagger\Phi}\Big)^2\Big\}+ \lambda_3 \Tr{\Phi^\dagger\tilde{\Phi}}\Tr{\tilde{\Phi}^\dagger\Phi}+ \lambda_4\Tr{\Phi^\dagger\Phi}\Tr{\Phi^\dagger\tilde{\Phi}+ \tilde{\Phi}^\dagger\Phi} \\
		&\	+\mu^{\prime}_1 (\chi^{\dagger}_L \Phi \chi_R +\chi^{\dagger}_R \Phi^{\dagger} \chi_L)+ \mu^{\prime}_2 (\chi^{\dagger}_L \tilde{\Phi} \chi_R +\chi^{\dagger}_R \tilde{\Phi}^{\dagger} \chi_L)+\rho_1 \bigg [ (\chi^{\dagger}_L \chi_L)^2 +(\chi^{\dagger}_R \chi_R)^2 \bigg ] \\
		&\ +\alpha_1 \Tr{\Phi^\dagger\Phi} \bigg [ (\chi^{\dagger}_L \chi_L) +(\chi^{\dagger}_R \chi_R) \bigg ] + \alpha_2 e^{i \delta} \bigg [ \Tr{ \tilde{\Phi} \Phi^\dagger} (\chi^{\dagger}_L \chi_L)+\Tr{\Phi \tilde{\Phi}^{\dagger}} (\chi^{\dagger}_R \chi_R) \bigg ] \\
		&\ + \alpha_2 e^{-i \delta} \bigg [ \Tr{ \tilde{\Phi}^{\dagger} \Phi} (\chi^{\dagger}_L \chi_L)+\Tr{\Phi^{\dagger} \tilde{\Phi}} (\chi^{\dagger}_R \chi_R) \bigg ]+\alpha_3 (\chi^{\dagger}_L \Phi \Phi^{\dagger} \chi_L + \chi^{\dagger}_R \Phi^{\dagger} \Phi \chi_R) \\
		&\		+\alpha_4 (\chi^{\dagger}_L \tilde{\Phi} \tilde{\Phi}^{\dagger} \chi_L + \chi^{\dagger}_R \tilde{\Phi}^{\dagger} \tilde{\Phi} \chi_R). \,
		\esp\label{appeneq2}\ee}           
For details of the minimisation of the scalar potential and resulting symmetry breaking, please refer to \cite{Bernard:2020cyi}. In the symmetry breaking pattern, the neutral component of the Higgs doublet $\chi_R$ acquires a vacuum expectation value (VEV) to break the gauge symmetry of the DLRM into that of the SM and then to the $U(1)$ of electromagnetism by the VEV of the neutral components of Higgs bidoublet $\Phi$:
	$$ SU(2)_L \times SU(2)_R \times U(1)_{B-L} \quad \underrightarrow{\langle
		\chi_R \rangle} \quad SU(2)_L\times U(1)_Y \quad \underrightarrow{\langle \Phi \rangle} \quad U(1)_{\rm em}.$$
The VEVs of the neutral components of the Higgs fields can be denoted as
\begin{center}
$\langle \Phi \rangle =
\begin{pmatrix}
\frac{k_1}{\sqrt{2}} & 0 \\
0 & \frac{k_2}{\sqrt{2}}\\
\end{pmatrix}~,~ \langle \chi_L \rangle =
\begin{pmatrix}
0 \\
\frac{v_L}{\sqrt{2}}
\end{pmatrix}~,~ \langle \chi_R \rangle =
\begin{pmatrix}
0 \\
\frac{v_R}{\sqrt{2}}
\end{pmatrix}$
\end{center}
where the VEV's $k_1, k_2$ satisfy the VEV of the SM namely, $v_{\rm SM} =\sqrt{k_1^2+k_2^2} \approx 246$~GeV. The spontaneous breaking of DLRM gauge symmetry down to $U(1)_{\rm em}$ results in two charged massive vector bosons $W_L, W_R$, two neutral massive bosons $Z_L, Z_R$ and a massless photon as expected. The details of the mass spectrum of gauge bosons are shown in appendix \ref{appen1}.

Light Dirac neutrino mass and charged lepton mass are given by 
\begin{equation}\label{eq4}
	M_{\nu}=\frac{1}{\sqrt{2}}(k_1h+k_2\widetilde{h}), M_{l}=\frac{1}{\sqrt{2}}(k_2h+k_1\widetilde{h})
	\end{equation}
where the family indices are suppressed. Without any loss of generality, we make
	use of rotation in the $SU(2)_L\times SU(2)_R$ space so that only one of the neutral components of the Higgs bidoublet acquires a large vacuum expectation value, $k_1\approx v_{\rm SM}$ and $k_2\approx0$. Under these assumptions, the Dirac neutrino mass matrix is 
	\begin{equation}
	M_{\nu}=\frac{1}{\sqrt{2}}(k_1h)
	\end{equation}
	while the charged lepton mass matrix is 
	\begin{equation}
	M_l=\frac{1}{\sqrt{2}}(k_1\widetilde{h})
	\end{equation}
Therefore, tiny sub-eV Dirac neutrino mass arises due to smallness of Yukawa coupling $h$ while charged lepton masses are generated by corresponding Yukawa coupling $\widetilde{h}$. The details of fermion-gauge boson couplings are shown in appendix \ref{appen2}. The details of the scalar mass spectrum is not derived here as we do not need them for our analysis and we refer to \cite{Bernard:2020cyi} for details of the same.

\section{$\Delta N_{\rm eff}$ in DLRM}
\label{sec2}
\begin{figure}[h!!]
\centering
\includegraphics[scale=0.5]{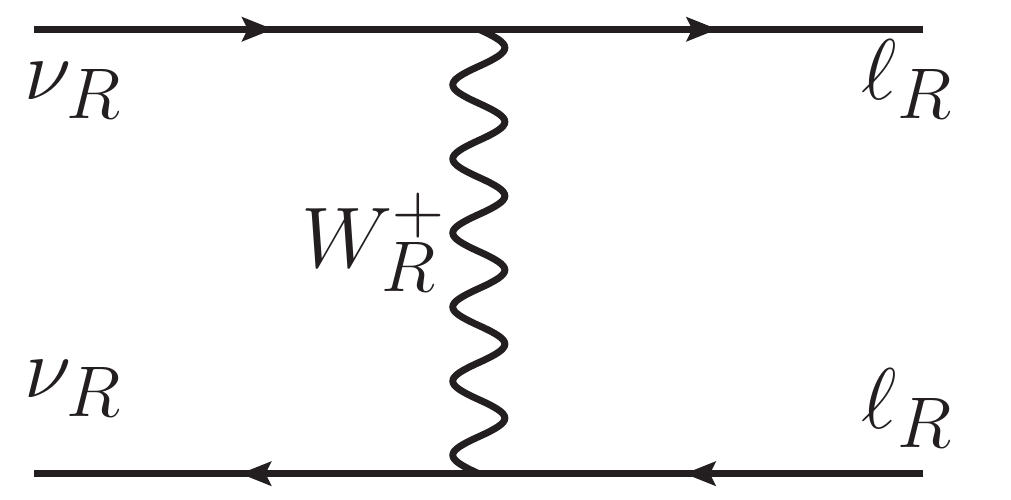}\,\,\, \includegraphics[scale=0.5]{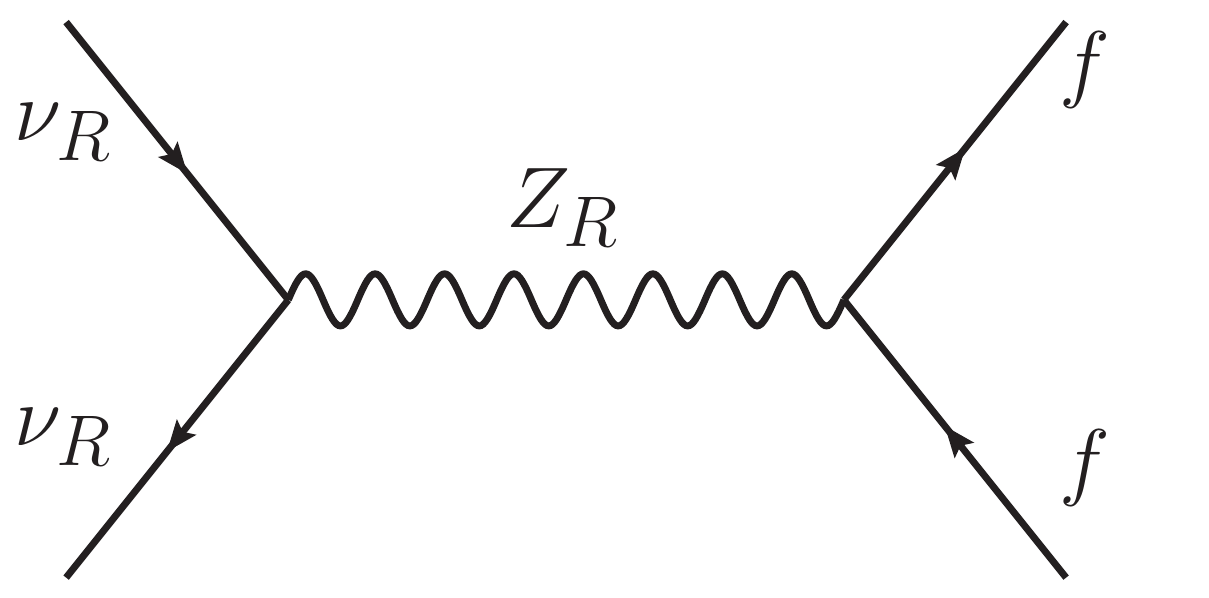}\\
\caption{Feynman diagrams of $\nu_R$ annihilation. Here $l_R \equiv e_R, \mu_R, \tau_R$, $f \equiv e_R, \mu_R, \tau_R, q_R$ with q being quark.}
\label{feyn}
\end{figure}

Effective number of relativistic degrees of freedom is defined as 
$$ N_{\rm eff} \equiv \frac{8}{7} \left( \frac{11}{4} \right)^{4/3} \left( \frac{\rho_{\rm rad} -\rho_{\gamma}}{\rho_{\gamma}} \right) $$
where $\rho_{\rm rad}=\rho_{\gamma} + \rho_{\nu}$ is the net radiation content of the universe. As mentioned earlier, the SM prediction is $N^{\rm SM}_{\rm eff}=3.045$ \cite{Mangano:2005cc, Grohs:2015tfy, deSalas:2016ztq} which is also consistent with the constraint from precision measurement of Z boson decay width at LEP $N_{\nu} = 2.984 \pm 0.008$ \cite{Tanabashi:2018oca}. Any deviation of $N_{\rm eff}$ from $N^{\rm SM}_{\rm eff}$ will therefore indicate the presence of additional relativistic species thermalised in the early universe. While these additional relativistic degrees of freedom can not fully thermalise with the SM bath through interactions mediated by Z boson due to strong LEP bound, they can thermalise via additional interactions or mediating particles not yet observed in direct search experiments. The right handed neutrinos in DLRM provides such an example. They can thermalise with the SM bath in the early universe due to the interactions mediated by right sector gauge bosons, as depicted by the Feynman diagrams shown in figure \ref{feyn}. We consider negligible mixing between left and right sector gauge bosons and hence ignore the contributions coming from processes like $\bar{\nu_R} \nu_L \rightarrow f \bar{f}, \bar{\nu_L} \nu_R \rightarrow f \bar{f}$. Additionally, the scalar mediated interactions are negligible due to tiny Dirac Yukawa couplings.

To estimate the contribution in $\Delta N_{\rm eff}$ we need to check the decoupling temperature of the right handed neutrinos. The decoupling occurs when the expansion rate of the universe becomes more than the interaction rate. Hence, the decoupling temperature can be calculated from the following equality
\begin{eqnarray}
\Gamma (T_{\nu_R}^{\rm d}) = H (T_{\nu_R}^{\rm d})
\label{eq7}
\end{eqnarray}
where $\Gamma(T)$ is the interaction rate and $H(T)$ is the expansion rate of the universe. The interaction rate can be written as 
\begin{eqnarray}
\Gamma (T)= n_{\nu_R} (T) \left\langle\sigma_{\rm Tot} v \right\rangle
\end{eqnarray}
where the number density $n_{\nu_R}$ for a relativistic neutrino can be written as 
\begin{eqnarray}
n_{\nu_R}(T)=\frac{3\,\, g_{\nu_R}}{4\,\, \pi^2}\,\, \zeta(3)\,\, T^3 
\end{eqnarray}
and the annihilation cross sections of right handed neutrinos are given in Appendix \ref{appen3}.

The expansion rate of the universe can be written as 
\begin{eqnarray}
H(T)=\sqrt{\frac{8\, \pi\, G_N \, \rho(T)}{3}}= \sqrt{\frac{4\pi^3G_N}{45}\left( g_{*}(T)+3\frac{7}{8}g_{\nu_R}\right)}T^2
\end{eqnarray}
where $g_{\nu_R}$ is the internal degrees of freedom for right-handed neutrinos. Thus, the contribution of $\nu_R$ to effective relativistic degrees of freedom can be estimated as
\begin{eqnarray}
\Delta N_{\rm eff} =N_{\rm eff}-N^{\rm SM}_{\rm eff} =N_{\nu_R}\left(\frac{T_{\nu_R}}{T_{\nu_L}}\right)^4= N_{\nu_R} \left(\frac{g_{*s}(T_{\nu_L}^{\rm d})}{g_{*s}(T_{\nu_R}^{\rm d})}\right)^{\frac{4}{3}}
\label{eqn:11}
\end{eqnarray}
where $\rm{N_{\nu_R}}$ represents the number of relativistic right-handed neutrinos, $g_{*}(T)$ corresponds to the relativistic degrees of freedom at temperature T, $g_{*s}(T)$ corresponds to the relativistic entropy degrees of freedom at temperature T \footnote{We use $g_{*}$ and $g_{*s}$ interchangeably, which is true in SM at high temperatures.} and ${\rm T^{d}_{\nu_R}\, \, , T^{d}_{\nu_L}}$ are the decoupling temperatures for ${\rm \nu_R}$ and ${\rm \nu_L}$ respectively. Thus, depending upon the decoupling temperature of $\nu_R$ and hence $g_{*}(T_{\nu_R}^{\rm d})$, the additional contribution to $\Delta N_{\rm eff}$ can be kept within experimental upper limits. Lower the strength of $\nu_R$ interaction with SM bath or higher the mediator mass of $\nu_R$-SM interactions, larger will be $g_{*}(T_{\nu_R}^{\rm d})$ and hence smaller will be $\Delta N_{\rm eff}$. Similar analysis for $U(1)_{B-L}$ extension of the SM can be found in \cite{Abazajian:2019oqj, FileviezPerez:2019cyn, Nanda:2019nqy, Han:2020oet} whereas some estimates in the context of radiative Dirac neutrino mass in LRSM were made in \cite{Borah:2016lrl, Borah:2017leo}.

\begin{figure}[h!!]
\centering
\includegraphics[scale=0.5]{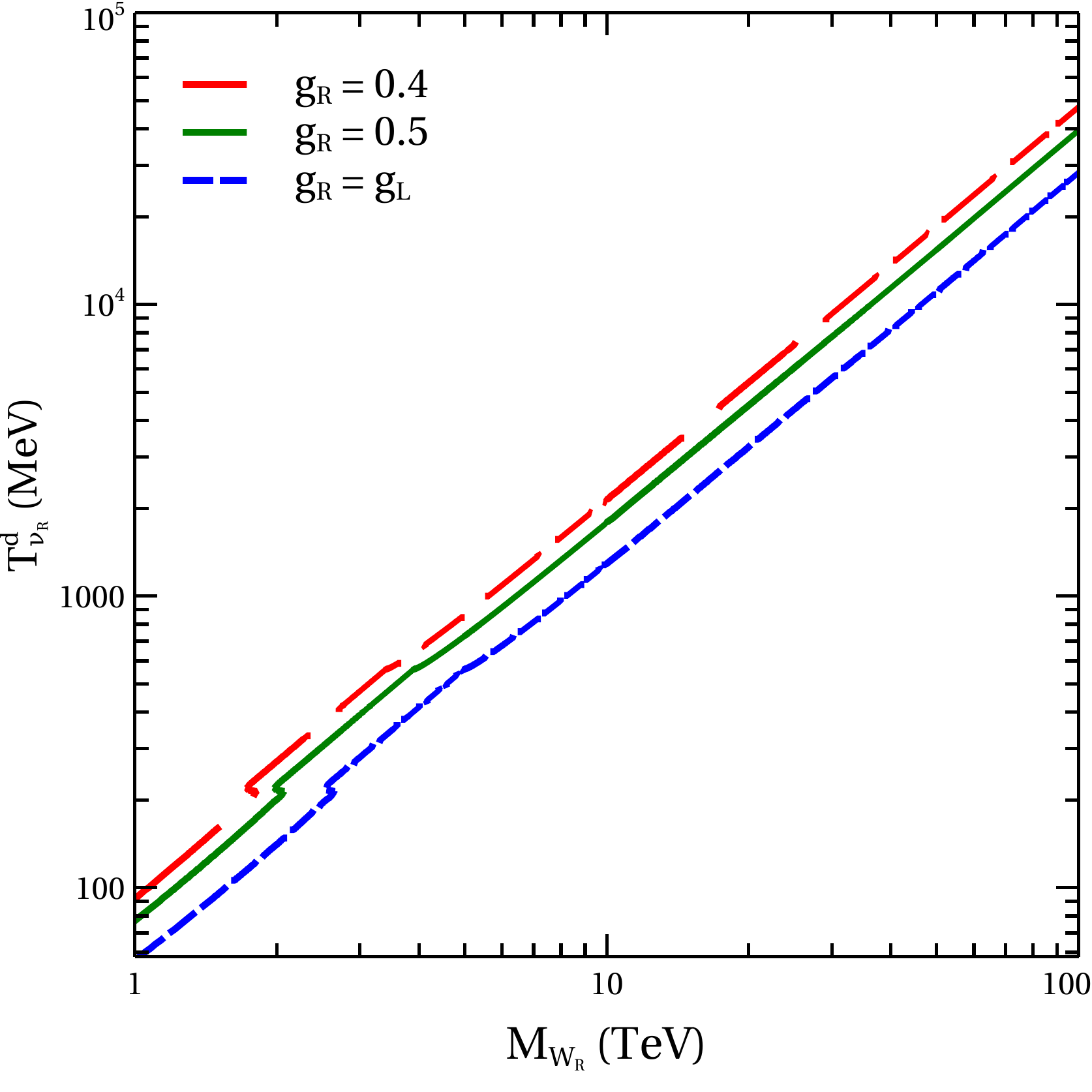}
\caption{Decoupling temperature of right handed neutrinos as a function of $W_R$ mass for different gauge couplings $g_R$.}
\label{fig2}
\end{figure}

\begin{figure}[h!!]
\centering
\includegraphics[scale=0.45]{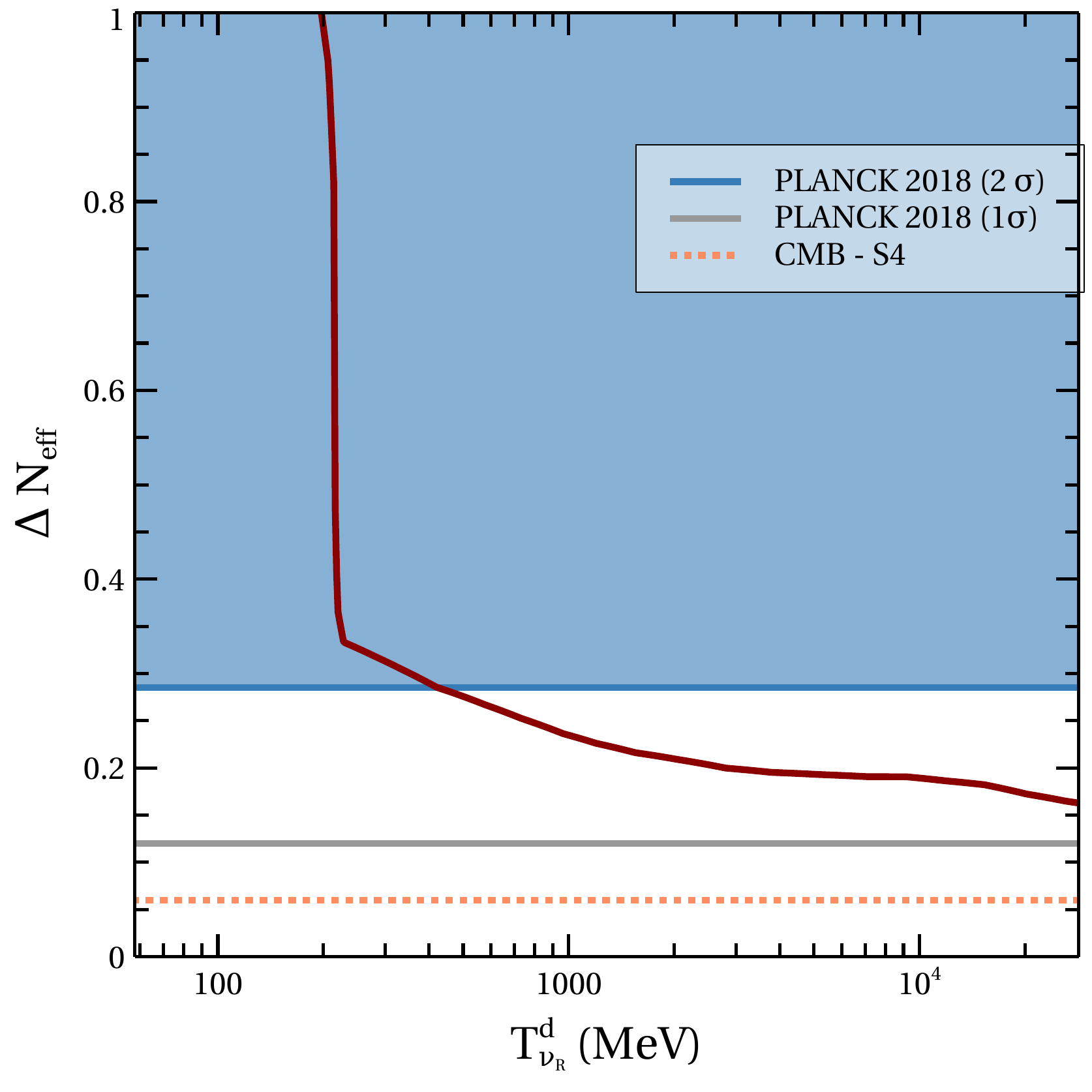}
\includegraphics[scale=0.45]{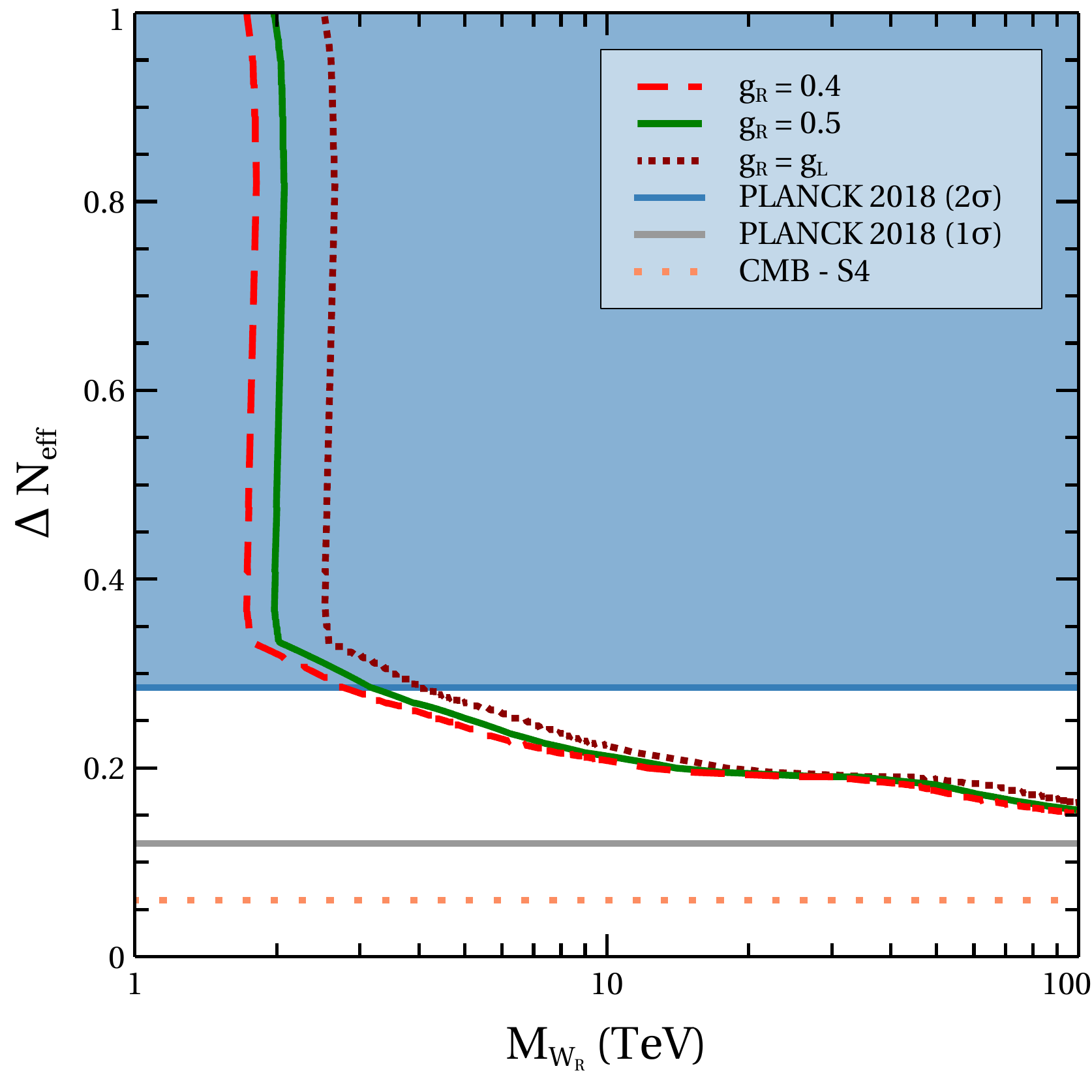}
\caption{$\Delta N_{\rm eff}$ as a function of decoupling temperature (left panel) and $W_R$ mass (right panel).}
\label{fig3}
\end{figure}

\begin{figure}[h!!]
\centering
\includegraphics[scale=0.5]{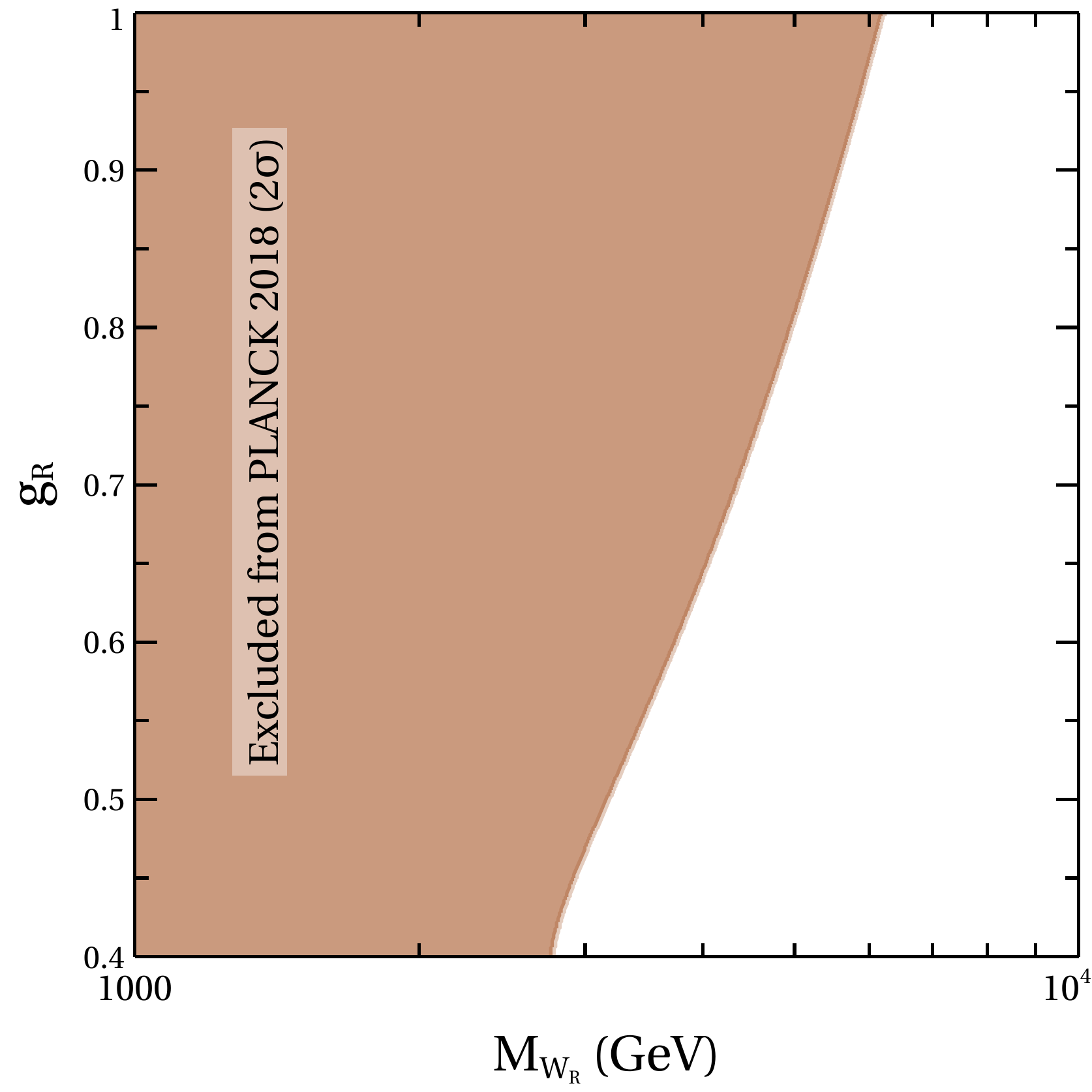}
\caption{Parameter space in $g_R-M_{W_R}$ plane from Planck 2018 $2\sigma$ constraints on $\Delta N_{\rm eff}$.}
\label{fig4}
\end{figure}

\begin{figure}[h!]
\centering
\includegraphics[width=0.5\textwidth]{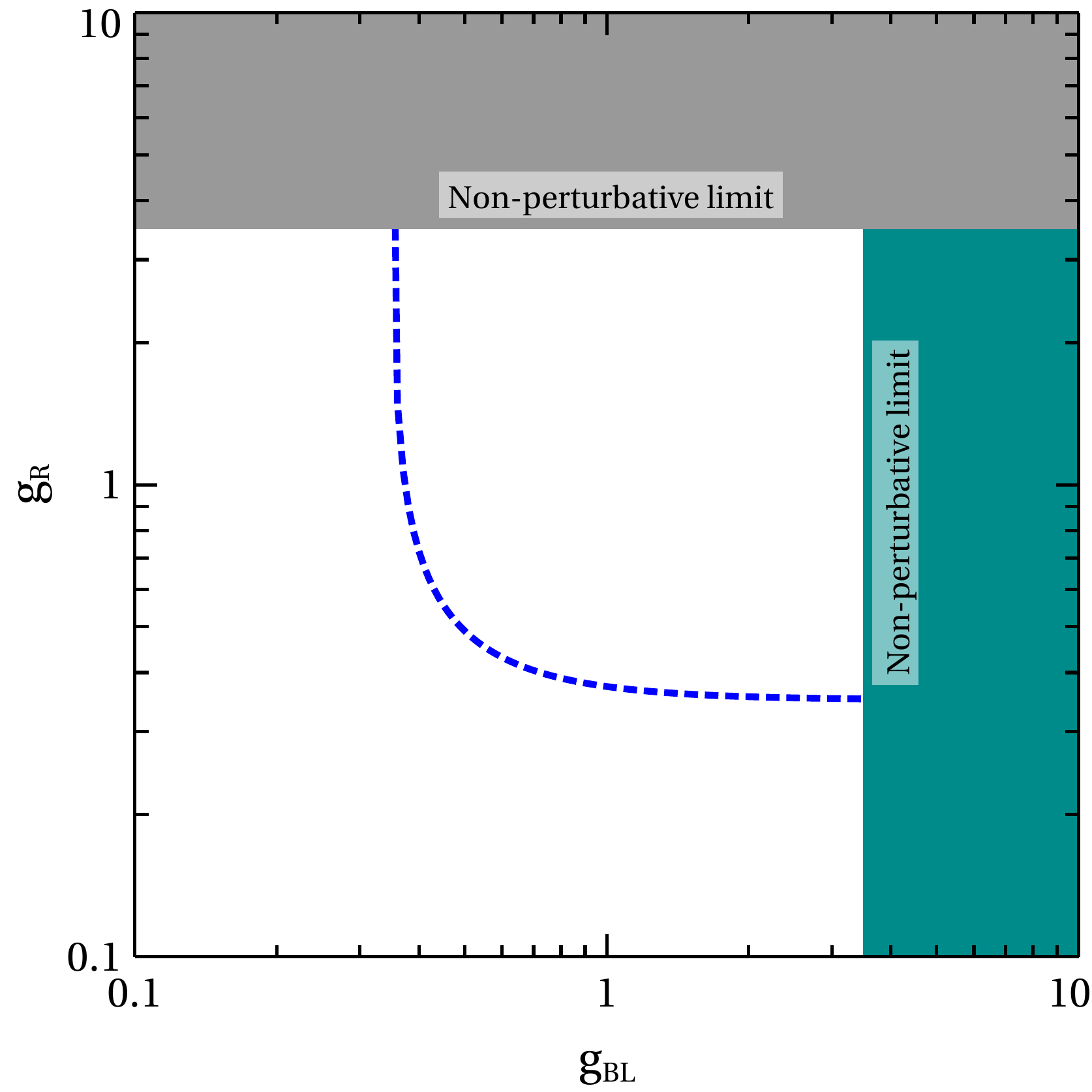}
\caption{Allowed values of $g_{BL}$ and $g_R$ which will reproduce the value $g_Y$ after the spontaneous breaking of ${\rm SU(2)_R \times U(1)_{B-L}}$ to the remaining $U(1)_{Y}$.} 
\label{fig1}
\end{figure}

\begin{figure}[h!]
\centering
\includegraphics[width=0.5\textwidth]{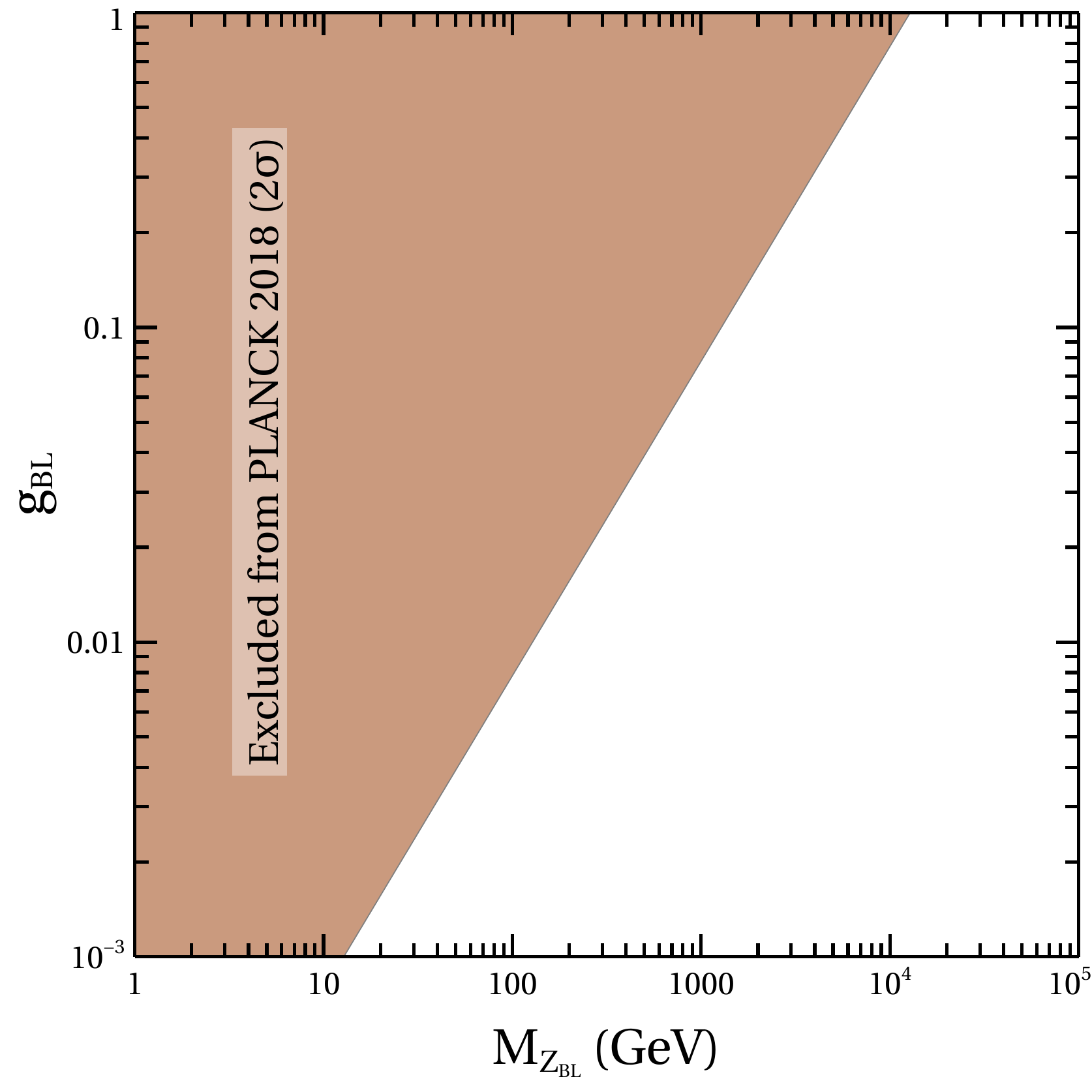}
\caption{Allowed $g_{BL}-M_{Z_{BL}}$ parameter space from Planck 2018 bound on $\Delta N_{\rm eff}$ at $2\sigma$ in minimal $U(1)_{B-L}$ gauge model with light Dirac neutrinos.} 
\label{fig1a}
\end{figure}

\begin{figure}[h!]
\centering
\includegraphics[width=0.47\textwidth]{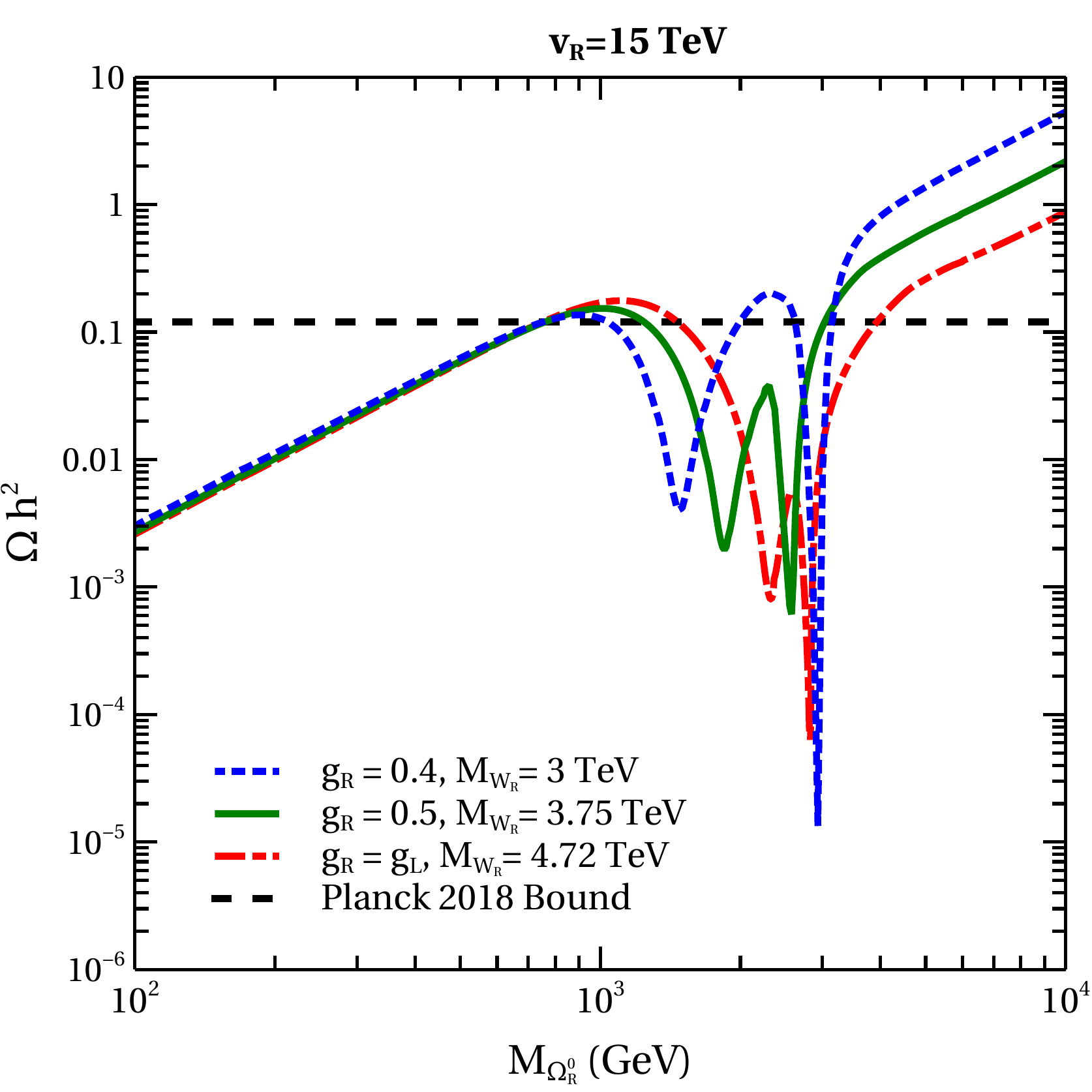}
\includegraphics[width=0.47\textwidth]{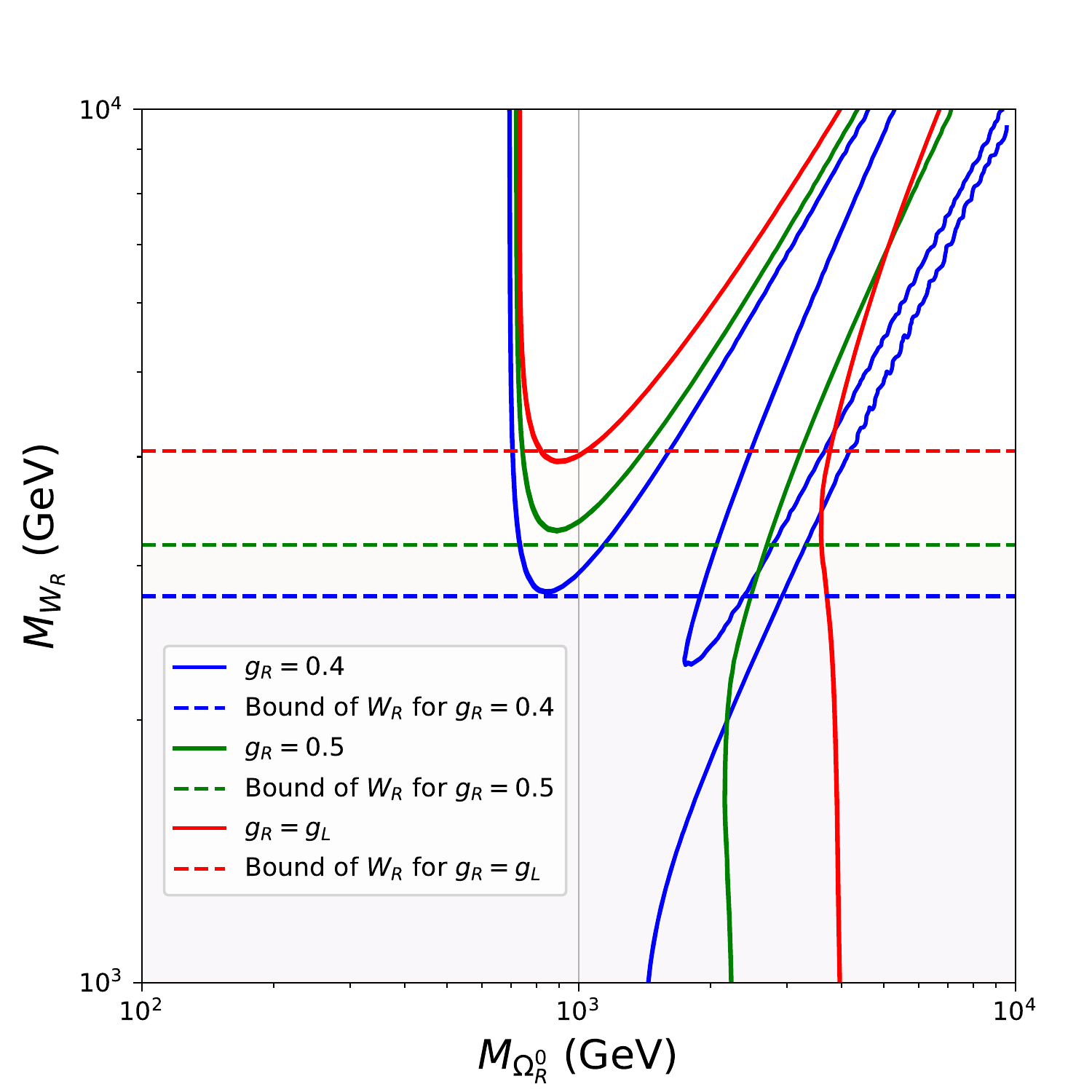}
\caption{Left panel: DM relic as a function of DM mass for different benchmark combinations of $g_R, W_R)$. Right panel: Parameter space satisfying relic abundance criteria of right handed fermion quintuplet dark matter in DLRM. The shaded regions are ruled out by Planck 2018 bound on $\Delta N_{\rm eff}$ at $2\sigma$ for respective values of $g_R$.} 
\label{fig5}
\end{figure}
\section{Dark Matter in DLRM}
\label{sec2a}
 The data from Planck experiment which restricts the effective relativistic degrees of freedom in our universe also reveal that more than 26\% of present universe's energy density is composed of a non-luminous and non-baryonic form of matter, known as dark matter. Apart from recent cosmology based experiments like Planck, there have been several astrophysical evidences for many decades suggesting the presence of DM \cite{Zwicky:1933gu,Rubin:1970zza,Clowe:2006eq}. In terms of density 
parameter $\Omega_{\rm DM}$ and $h = \text{Hubble Parameter}/(100 \;\text{km} ~\text{s}^{-1} 
\text{Mpc}^{-1})$, the present DM abundance is conventionally reported as \cite{Aghanim:2018eyx}:
$\Omega_{\text{DM}} h^2 = 0.120\pm 0.001$
at 68\% CL. Given that none of the SM particles can be a viable DM candidate, several BSM proposals have been put forward among which the weakly interacting massive particle (WIMP) paradigm is the most popular one. In this framework, a DM particle having masses and interactions similar to those around the electroweak scale gives rise to the observed relic after thermal freeze-out, a remarkable coincidence often referred to as the {\it WIMP Miracle} \cite{Kolb:1990vq}.

The minimal DLRM discussed above does not have a stable DM candidate. One can however, minimally extend the model by including additional scalar or fermionic multiplets in the spirit of minimal dark matter scenario \cite{Cirelli:2005uq,Garcia-Cely:2015dda,Cirelli:2015bda}. In these models, the dark matter candidate is stabilised either by a $Z_2 = (-1)^{B-L}$ subgroup of the $U(1)_{B-L}$ gauge symmetry or due to an accidental symmetry at the renormalisable level due to the absence of any renormalisable operator leading to dark matter decay. Such minimal dark matter scenario in LRSM has been studied recently by the authors of \cite{Heeck:2015qra,Garcia-Cely:2015quu}.  Some more recent works on DM in LRSM can be found in \cite{Borah:2016hqn, Borah:2016ees, Berlin:2016eem, Borah:2016lrl, Borah:2017leo, Borah:2017xgm, Dev:2016qbd, Dev:2016xcp, Dev:2016qeb, Borah:2017hgt, Ko:2015uma}. Unlike in triplet LRSM where $SU(2)_R \times U(1)_{B-L}$ gauge symmetry is spontaneously broken by scalar triplet with even $(B-L)$ charge $(\pm 2)$, in DLRM the same happens due to scalar doublet with odd  $(B-L)$ charge $(\pm 1)$. Thus, there is no stabilising symmetry like $Z_2 = (-1)^{B-L}$ in DLRM to stabilise DM. Therefore some DM candidates like fermion triplet, fermion bidoublet, scalar doublet discussed in the context of triplet LRSM \cite{Heeck:2015qra,Garcia-Cely:2015quu} are no longer stable in DLRM due to the presence of renormalisable interactions with lighter fields. We therefore consider the option of larger fermion multiplet as DM, and the minimal scenario is to consider a real fermion quintuplet of $B-L$ charge 0. Since we want to constrain the right sector gauge bosons from cosmology bound on $\Delta N_{\rm eff}$, we particularly focus on right handed fermion quintuplet DM whose relic abundance depends upon the strength of its annihilation through right sector gauge bosons.

In the pure left-right symmetric setup, one has to introduce a pair of left and right handed fermion quintuplets (having same mass) which can be written in component form as 
\begin{equation}
\Omega_{L}=\begin{pmatrix}
  \Omega^{++}_{L}  \\
   \Omega^{+}_{L} \\
     \Omega^{0}_{L} \\
       \Omega^{-}_{L} \\
         \Omega^{--}_{L} \\
 \end{pmatrix},  \;\;
\Omega_{R}=\begin{pmatrix}
  \Omega^{++}_{R}  \\
   \Omega^{+}_{R} \\
     \Omega^{0}_{R} \\
       \Omega^{-}_{R} \\
         \Omega^{--}_{R} \\
 \end{pmatrix}.
\end{equation}
Since we are discussing a general scenario with $g_L \neq g_R$, we consider the left fermion quintuplet to be very heavy and decoupled from the low energy phenomenology. Even in the pure left-right symmetric limit $g_L = g_R$, one can make the left quintuplet decouple from the low energy phenomenology by introducing a parity odd scalar singlet whose non-zero VEV at a very high scale splits the right and left fermion masses. Such proposals where the left-right discrete symmetry or parity gets broken spontaneously before $SU(2)_R \times U(1)_{B-L}$ gauge symmetry were put forward long ago in \cite{Chang:1983fu, Chang:1984uy, Chang:1984qr}. While all the components of fermion multiplet have same tree level masses, at radiative level, there arises a mass splitting between charged (with Q) and neutral components given by \cite{Heeck:2015qra,Garcia-Cely:2015quu},
\begin{align}
M_{\Omega_R^Q}-M_{\Omega_R^0} &\simeq \frac{\alpha_2}{4\pi} \frac{g_R^2}{g_L^2} M Q^2\left[ f(r_{W_R}) - c_M^2 f(r_{Z_R})-s_W^2 s_M^2 f(r_{Z_L})- c_W^2 s_M^2 f(r_\gamma)\right],
\end{align}
where $s_M = \sin{\theta_M} \equiv \tan{\theta_W} \frac{g_L}{g_R}, s_W = \sin{\theta_W}, r_X = M_X/M$ and $$ f(r) \equiv 2 \int^1_0 dx (1+x) \log{[x^2+(1-x)x^2]} ~.$$ The bare mass of the multiplet is denoted by $M$. Here the one loop self-energy corrections through mediations of gauge bosons are presented within the square bracket of the second expression. Due to such tiny one loop mass splitting, the next to lightest component of each DM multiplet can be thermally accessible during the dark matter freeze-out and hence the coannihilation effects play a crucial role \cite{Griest:1990kh}. While in triplet LRSM, there exists the possibility that for $M \gg M_{W_R}$, the neutral component of the multiplet can become heavier compared to the charged components, such possibilities do not arise in DLRM. This was noted in \cite{Ko:2015uma} which we also confirm. Apart from its role in enhancing DM coannihilations, such mass splitting, if small enough, may also induce inelastic DM-nucleon scattering mediated by $W_R$ bosons. However, for our region of interest, such tiny mass splitting does not arise. This ensures that $W_R$ mediated DM-nucleon scattering occurs only at one loop level and hence remain suppressed. Spin-independent elastic DM nucleon scattering mediated by $Z_R$ at tree level remains absent due to vanishing $B-L$ charge of fermion quintuplet discussed in this work. Thus, the DM phenomenology of right handed fermion quintuplet is mainly governed by its gauge interactions which are given by  \cite{Ko:2015uma}
\begin{align}
\mathcal{L}_{\Omega_R} & \supset -s_W s_M g_R Q \overline{\Omega^Q_R} Z^{\mu}_L \gamma_{\mu} \Omega^Q_R + c_M g_R Q \overline{\Omega^Q_R} Z^{\mu}_R \gamma_{\mu} \Omega^Q_R \nonumber \\
& +c_W s_M g_R Q \overline{\Omega^Q_R} A^{\mu} \gamma_{\mu} \Omega^Q_R + \frac{g_R}{\sqrt{2}} \left (c_{Q} \overline{\Omega^{Q+1}_R} W^{\mu}_R \gamma_{\mu} \Omega^Q_R + {\rm h.c.} \right)
\end{align}
where $c_{Q} = \sqrt{(2+Q+1)(2-Q)}$ and $Q$ is the electromagnetic charge of the quintuplet component.

\section{Results and Discussion}
\label{sec3}
Using the recipe discussed in previous section, we first calculate the decoupling temperature of right handed neutrinos from the thermal bath for different values of $W_R, Z_R$ mass and gauge coupling $g_R$. The variation of decoupling temperature with $W_R$ mass for different values of $g_R$ is shown in figure \ref{fig2}. Although both $W_R$ and $Z_R$ masses play role in right handed neutrino interactions with the thermal bath, we show the variation of decoupling temperature as well as other physical quantities only in terms of $W_R$ mass. This is due to the fact that $Z_R$ mass typically depends upon $W_R$ mass and is heavier than it, similar to $Z$ and $W$ masses of the SM. Also, we are not restricting ourselves to pure left-right symmetric limit $g_R=g_L$ and considering different values of $g_R$ as well. Decoupling temperature rises for lower values of gauge coupling as well as higher values of $W_R$ mass as seen from figure \ref{fig2} which is expected as the corresponding rate of interactions decreases. Typically the interaction cross section $ \langle \sigma v \rangle$ of right handed neutrinos depends upon temperature as $g^4_R T^2/M^4_{W_R}$ and hence the rate of interaction is $\Gamma (T) \propto g^4_R T^5/M^4_{W_R}$. When this is compared with the Hubble parameter $H(T) \propto T^2$ as in equation \eqref{eq7} to find the decoupling temperature ${\rm T_{\nu_R}^d}$, it is clear that with increase in $W_R$ mass, the decoupling temperature will also rise. Similarly, decrease in $g_R$ will lead to an increase in decoupling temperature for same value of $M_{W_R}$. This conclusion reached from approximate analytical formulas for cross section and Hubble parameter also agrees with our numerical results shown in figure \ref{fig2}.

We then show the contribution to $\Delta N_{\rm eff}$ in figure \ref{fig3} as functions of decoupling temperature as well as $W_R$ mass. The dependance of $\Delta N_{\rm eff}$ on ${\rm T_{\nu_R}^d}$ in the left panel of figure \ref{fig3} can be understood from equation \eqref{eqn:11}. The total number of relativistic degrees of freedom at decoupling temperature $g_{*s}(T_{\nu_R}^{\rm d})$ increases with the increasing ${\rm T_{\nu}^d}$ which thereby decreases the value of $\Delta N_{\rm eff}$. However, at some point $g_{*s}$ it will reach its maximum value and $\Delta N_{\rm eff}$ becomes almost constant, as seen from the plateau region on bottom left corner of left panel plot in figure \ref{fig3}. Since $T^d_{\nu_R}$ is being varied independently and corresponding $g_{*s}(T_{\nu_R}^{\rm d})$ is used to find $\Delta N_{\rm eff}$ using equation \eqref{eqn:11}, the behaviour of this plot does not depend upon $g_R, M_{W_R}$. To show the dependence of $\Delta N_{\rm eff}$ on such model parameters, we have made the plot shown in right panel of figure \ref{fig3}. The right panel of figure \ref{fig3} represents the dependance of $\Delta N_{\rm eff}$ on ${\rm M_{W_R}}$ for three benchmark values of $g_R$. One can see that the contribution to the $\Delta N_{\rm eff}$ decreases with increasing $M_{W_R}$. This is because, for higher values of $M_{W_R}$, $\nu_R$ will decouple at some higher temperature and the contribution to the $\Delta N_{\rm eff}$ will become smaller. Along with the Planck 2018 bound mentioned earlier, we also show the CMB-S4 sensitivity \cite{Abitbol:2019nhf} as well as the Planck 2018 $1\sigma$ limit while the latter is same as SPT-3G sensitivity \cite{Benson:2014qhw}. Clearly, Planck 2018 bound at $2\sigma$ CL itself rules out $W_R$ mass below 4.06 TeV with gauge coupling $g_R = g_L$. On the other hand, future probe will be able to either confirm or rule out the model, even for very high $W_R$ masses, out of reach of direct search experiments. Finally we show the final parameter space in $g_R-M_{W_R}$ plane after applying Planck 2018 $2\sigma$ constraints in figure \ref{fig4}.

Unlike in \cite{Abazajian:2019oqj, FileviezPerez:2019cyn, Nanda:2019nqy, Han:2020oet} where similar constraints on $U(1)_{B-L}$ gauge boson was obtained, the crucial difference in DLRM is that here one can not tune the gauge couplings for a particular value of gauge boson mass in order to suppress the contribution to $\Delta N_{\rm eff}$. This is because the gauge couplings of $SU(2)_R$ and $U(1)_{B-L}$ are not arbitrary but related to the gauge coupling of $U(1)_Y$ (at the scale of left-right symmetry breaking) as 
\begin{equation}
\frac{1}{g^2_Y} = \frac{1}{g^2_R} +\frac{1}{g^2_{BL}} 
\end{equation}
Since $g_Y$ is known, one can not change $g_R, g_{BL}$ arbitrarily within their perturbative limits\footnote{For discussion related to perturbativity constraints on similar models, please refer to \cite{Chauhan:2018uuy} and references therein.}. We show the allowed region of these two gauge couplings in figure \ref{fig1}. While we still have a large region within perturbative limits, we have chosen $g_R$ to be either equal to $g_L$ or smaller while keeping $g_{BL}$ also below order one for our benchmark analysis. For $g_R > g_R$ the Planck bound becomes even more stringent, as we found in the scan plot shown in $g_R-M_{W_R}$ plane in figure \ref{fig4}. Thus, compared to $U(1)_{B-L}$ or other Abelian gauge models of Dirac neutrinos, DLRM is much more constrained. For a comparison we show the parameter space for gauged $U(1)_{B-L}$ model with Dirac neutrinos in figure \ref{fig1a}. This is a minimal gauged $B-L$ model where there are three right handed neutrinos having $B-L$ charge -1 each apart from the SM fermion content and neutrinos get sub-eV Dirac mass by virtue of their tiny couplings with SM Higgs. Since the $B-L$ gauge coupling is a free parameter and not related to SM gauge couplings in this model, one can tune the gauge coupling arbitrarily to evade the stringent Planck bound on $\Delta N_{\rm eff}$ as can be seen from figure \ref{fig1a}.

To check the impact of these constraints on DM parameter space, we then calculate relic of right fermion quintuplet DM. For DM relic calculation, we first implement the model in \texttt{SARAH} \cite{Staub:2013tta} and then feed the model files into \texttt{micrOMEGAs} \cite{Belanger:2013oya} for relic calculations. We then consider three benchmark combinations of $(g_R, W_R)$ while keeping the scale of left-right gauge symmetry breaking $v_R$ fixed. The resulting variation of DM relic as a function of DM mass is shown in left panel plot of figure \ref{fig5}. The resonance corresponding to $W_R, Z_R$ masses are clearly visible in this plot. Unlike in quintuplet DM scenario in triplet LRSM \cite{Garcia-Cely:2015quu}, here the two resonances are quite close to each other due to smaller ratio of $Z_R$ to $W_R$ mass in DLRM. The results shown in left panel plot of figure \ref{fig5} also agree with that shown in \cite{Ko:2015uma}. We then scan the parameter space of $W_R$, $\Omega^0_R$ masses and show the region satisfying correct DM relic in right panel plot of figure \ref{fig5}. Multiple allowed values of DM mass for a fixed $W_R$ mass are arising due to annihilation and coannihilations of $\Omega^0_R, \Omega^{\pm}_R, \Omega^{\pm \pm}_R$ mediated by $W_R, Z_R$ bosons where $Z_R$ is slightly heavier than $W_R$ ($M_{Z_R} \approx 1.2 M_{W_R}$ in pure left-right symmetric limit $g_R = g_L$). Further, red and green solid lines on the right panel plot of figure \ref{fig5} has shows the possibility of having three different values of DM mass with correct relic abundance for a fixed $W_R$ mass. The same behaviour is also seen on the left panel plot of figure \ref{fig5} with fixed $W_R$ mass where the red and green lines satisfy relic at three different values of DM masses. On the other hand, the blue dashed line on the left panel plot of figure \ref{fig5} satisfies correct relic abundance for five different values of DM masses, a feature which is also depicted by the blue solid line in the scan plot (right panel of figure \ref{fig5}). We also apply the corresponding bounds on $W_R$ mass from Planck constraints on $\Delta N_{\rm eff}$ at $2\sigma$ CL as horizontal shaded lines so that the region below the respective lines are disallowed. Clearly, some part of the DM parameter space on the right side of the scan plot (right panel of figure \ref{fig5}) gets disfavoured for all three values of $g_R$ by $\Delta N_{\rm eff}$ bounds as the lower region of the dashed lines are excluded from Planck constraints on $\Delta N_{\rm eff}$ at $2\sigma$ CL. In the parabolic part of the parameter space in the scan plot, a small part of the parameter space for $g_R=g_L$ case is disfavoured by $\Delta N_{\rm eff}$ bounds while the parabolic lines corresponding to $g_R \neq g_L$ still remains allowed from respective $\Delta N_{\rm eff}$. We do not show other existing bounds on $W_R$ mass from flavour or LHC data as they are either equally or less strong compared to the bounds derived here.
\begin{figure}[h!]
\centering
\includegraphics[width=0.47\textwidth]{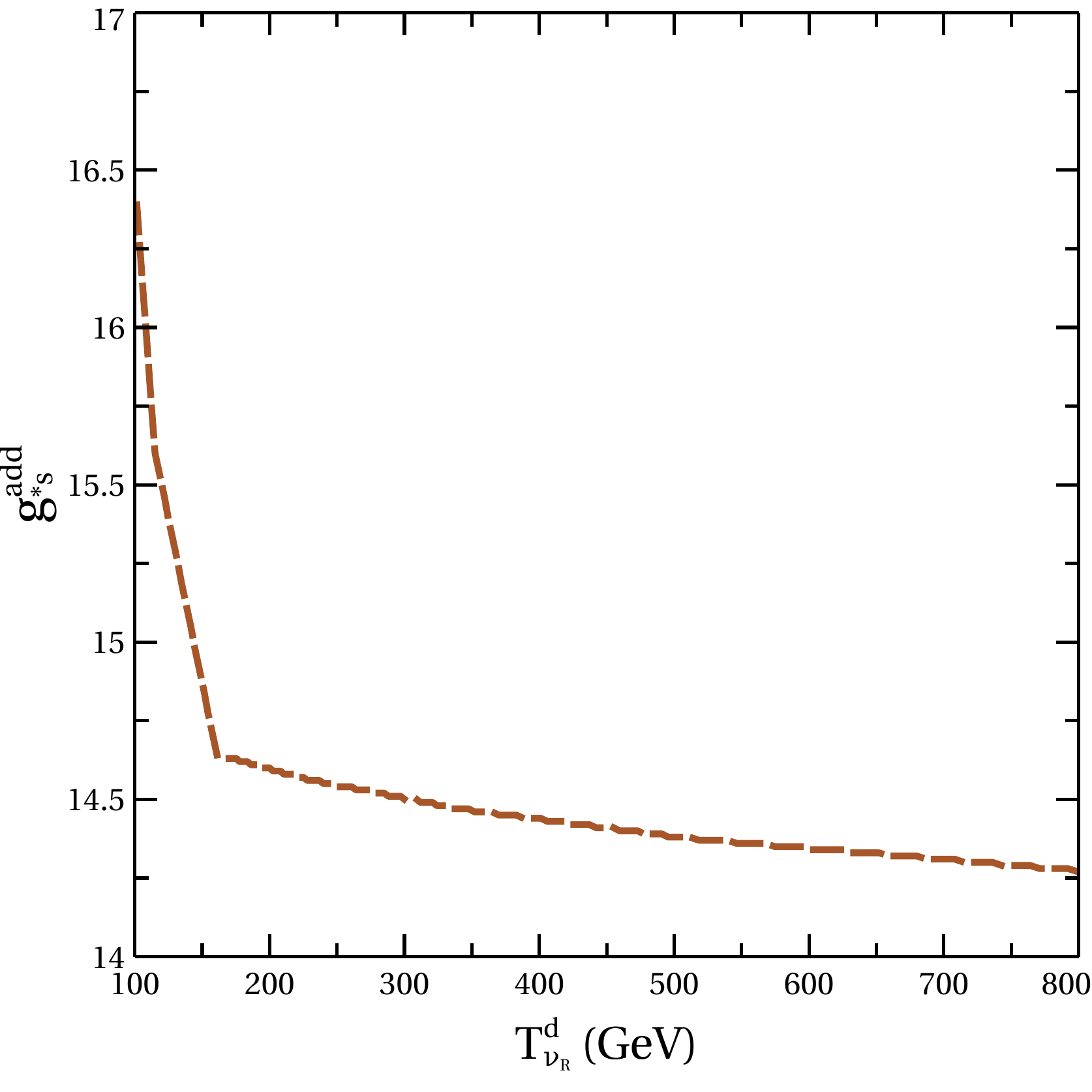}
\includegraphics[width=0.47\textwidth]{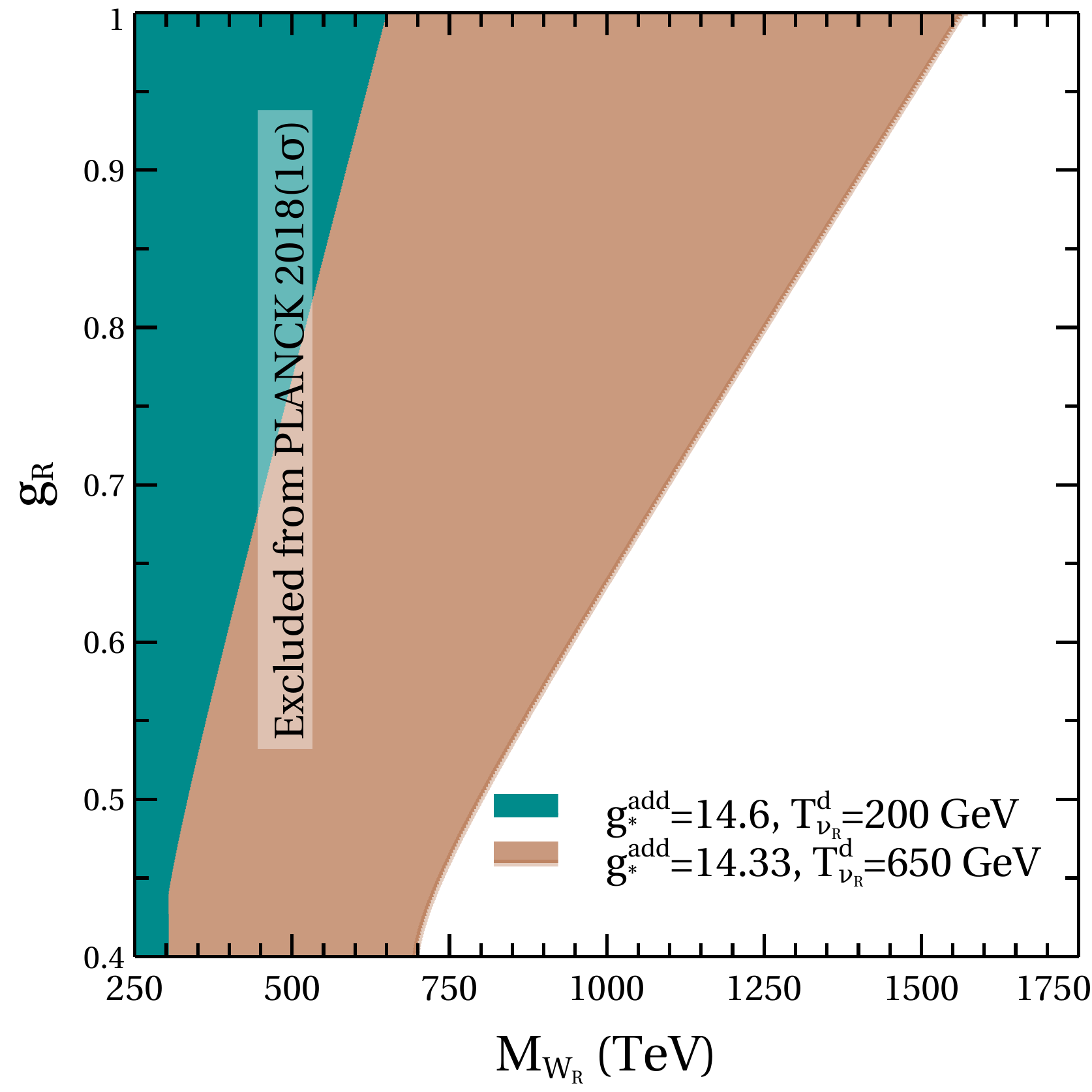}
\caption{Left panel: Additional relativistic degrees of freedom required to bring $N_{\rm eff}$ within Planck 2018 $1\sigma$ bound $N_{\rm eff} = 2.99 \pm 0.17$. Right panel: Allowed parameter space in $g_R-M_{W_R}$ plane from Planck 2018 $1\sigma$ bound $N_{\rm eff} = 2.99 \pm 0.17$ after considering additional relativistic degrees of freedom.} 
\label{fig6}
\end{figure}

\section{Conclusion}
\label{sec4}
We have studied the minimal left-right symmetric model with Higgs doublets, known as doublet left-right model where the left-right symmetry is broken spontaneously by Higgs doublet instead of Higgs triplets in LRSM with type I plus type II seesaw for light neutrino masses. In the minimal DLRM, light neutrino mass can be realised only through the Higgs bidoublet with tiny coupling to neutrinos leading to sub-eV Dirac neutrino mass. Due to $SU(2)_R \times U(1)_{B-L}$ gauge interactions, the right handed neutrinos can thermalise in the early universe thereby contributing to the effective relativistic degrees of freedom $N_{\rm eff}$ which is tightly constrained by CMB measurements. We constrain the scale of left-right symmetry from the requirement of satisfying Planck 2018 bound on $\Delta N_{\rm eff}$ not only for pure left-right symmetric limit $g_R = g_L$ but also for other values of $g_R$. While $g_R, g_{BL}$ can not be chosen arbitrarily in LRSM due to their relation with $U(1)_Y$ coupling of SM, the parameter space gets tightly constrained. For $g_R = g_L$, Planck 2018 bound at $2\sigma$ CL can rule out $W_R$ mass all the way upto 4.06 TeV which is as competitive as the existing collider bounds on $W_R$ from dijet resonance searches. For example, ATLAS dijet resonance search rules out such additional charged vector boson mass upto 4 TeV \cite{Aad:2019hjw} while similar analysis by CMS rules out upto 3.6 TeV for SM like gauge couplings \cite{Sirunyan:2019vgj}. Our conclusions also agree with the recent model independent calculations \cite{1794147} where, considering four fermion interactions of right handed neutrinos and their contribution to $N_{\rm eff}$, the authors constrained the interaction strength to be $10^{-5}-10^{-3}$ times the usual Fermi coupling constant. This bound we derive here that is, $M_{W_R} > 4.06$ TeV can be made weaker for smaller values of $g_R$. For example, in case of $g_R=0.5$, we get $M_{W_R} > 3.17$ TeV and $g_R=0.4$ leads to $M_{W_R} > 2.77$ TeV. We also make a comparison with similar constraints derived in $U(1)_{B-L}$ gauge model with light Dirac neutrinos where due to the freedom in choosing the gauge coupling arbitrarily, one can have much lighter $B-L$ gauge boson as well. However the choices of $g_R$ as well as $g_{BL}$ are not arbitrary in DLRM due to their non-trivial connection to $g_Y$ of standard model. While we perform our analysis without explicitly solving the Boltzmann equations, the conclusions do not change much as long as the decoupling temperature of right handed neutrinos remain much higher compared to that of active neutrinos. This was also noted by the authors of \cite{1794147}, and our parameter space is precisely confined to this regime.

We also show the impact of $\Delta N_{\rm eff}$ constraints on dark matter parameter space in DLRM. While DLRM does not have a dark matter candidate on its own, we incorporate the presence of an additional fermion quintuplet DM in the minimal DM spirit. Since such a real fermion quintuplet does not have any renormalisable coupling with other fermions or scalars of DLRM, the relic abundance of DM, the neutral component of right-handed fermion quintuplet, depends crucially on its annihilation and coannihilation mediated by $W_R, Z_R$ gauge bosons. We constrain the parameter space satisfying correct DM relic by using the respective $\Delta N_{\rm eff}$ bounds for different $g_R$. We find available parameter space satisfying correct DM relic even after applying Planck 2018 bound on $\Delta N_{\rm eff}$ at $2\sigma$ CL.

We also compare our results in view of more stringent Planck 2018 $1\sigma$ bound $N_{\rm eff} = 2.99 \pm 0.17$ which rules out all the parameter space if we assume only SM plus three right handed neutrinos to be contributing to the relativistic degrees of freedom (DOF) below the scale of left-right symmetry breaking. However, if there are more relativistic degrees of freedom due to the presence of light physical fields resulting from the scalar and fermion multiplets, one can satisfy the Planck 2018 $1\sigma$ bound as well. We show the required additional DOF in left panel of figure \ref{fig6} as a function of right handed neutrino decoupling temperature. All the points on the dashed line in left panel of figure \ref{fig6} gives rise to $\Delta N_{\rm eff} = 0.12$ so that the points below this line are ruled out. These DOF can arise from DLRM with right handed fermion quintuplet DM. For example, the scalar bidoublet has four physical DOF apart from the SM Higgs while the pair of Higgs doublets can give rise to five more physical DOF. Similarly, right handed fermion quintuplet DM has approximately ten DOF. Thus, one can have 19 additional DOF in DLRM with right handed fermion quintuplet DM. Similar ways of avoiding such strict cosmological bounds on $\Delta N_{\rm eff}$ have also been discussed in the recent work \cite{1794147}. On the right panel of figure \ref{fig6}, we show the allowed parameter space on $g_R-M_{W_R}$ plane from Planck 2018 $1\sigma$ bound after considering the required additional relativistic DOF at decoupling temperatures of 200 GeV and 650 GeV respectively. Since decoupling temperature of 200 GeV can be achieved with lighter $W_R$ gauge boson, therefore we get more allowed parameter space compared to the case with higher decoupling temperature of right handed neutrinos. However, presence of such additional light degrees of freedom will face stringent tests from collider as well as flavour physics constraints. Additionally, possible UV completion of this minimal model in order to explain the tiny origin of Dirac neutrino mass naturally \cite{Babu:1988yq, Ma:1989tz, Ma:2016mwh, Borah:2016lrl, Borah:2016hqn, Borah:2017leo, Ma:2017kgb, Chavez:2019yal} may involve additional interaction portals of right handed neutrinos to thermalise with the standard bath. While the strength of these portals can be tuned with more freedom compared to the gauge portals discussed in this work and hence are unlikely to put stronger constraints than what we obtain here, it will be interesting to study the details of such scenarios, specially in the context of dark matter and collider phenomenology. A more detailed investigation of such scenarios are left for future studies.

\acknowledgements
DB acknowledges the support from Early Career Research Award from DST-SERB, Government of India (reference number: ECR/2017/001873). CM wants to thank Supriya Senapati for useful discussion.

\appendix

\section{Physical Masses of Gauge Bosons}
\label{appen1}
Covariant derivatives of the scalar fields in DLRM can be written as
\begin{eqnarray}
D_{\mu} \Phi &=& \partial_{\mu} \Phi - i \frac{g_L}{2} (\sigma. W_{L\mu})\Phi + i \frac{g_R}{2} \Phi (\sigma. W_{R\mu}) \nonumber \\
 D_{\mu} \chi_L &=& \partial_{\mu}\, \chi_{L} - i \frac{g_L}{2} (\sigma. W_{L\mu})\chi_L - i \,\, g_{BL}\,Q_{BL} B^\prime_{\mu} \chi_{L} \nonumber \\
 D_{\mu} \chi_R &=& \partial_{\mu}\, \chi_{R} - i \frac{g_R}{2} (\sigma. W_{R\mu})\chi_R - i \,\, g_{BL}\,Q_{BL} B^\prime_{\mu}\chi_R
\end{eqnarray}
where
\begin{equation}
\sigma.W_{L/R \mu}=\left(
\begin{array}{cc}
 W_{L/R\mu}^3 &\sqrt{2} W_{L/R \mu}^+ \\
 \sqrt{2} W_{L/R \mu}^- & - W_{L/R\mu}^3 \\
\end{array}
\right).
\end{equation}
The corresponding kinetic Lagrangian of scalar fields are
\begin{eqnarray}
\mathcal{L}_{scalar}= Tr\big[\big(D_{\mu}\Phi\big)^\dagger \big(D_{\mu}\Phi\big)\big] + \big(D_{\mu}\chi_L\big)^\dagger \big(D_{\mu}\chi_L\big) + \big(D_{\mu}\chi_R\big)^\dagger \big(D_{\mu}\chi_R\big).
\end{eqnarray}

Considering the scalar vevs as,
\begin{center}
$\langle \Phi \rangle =
\begin{pmatrix}
\frac{k_1}{\sqrt{2}} & 0 \\
0 & \frac{k_2}{\sqrt{2}}\\
\end{pmatrix}~,~ \langle \chi_L \rangle =
\begin{pmatrix}
0 \\
\frac{v_L}{\sqrt{2}}
\end{pmatrix}~,~ \langle \chi_R \rangle =
\begin{pmatrix}
0 \\
\frac{v_R}{\sqrt{2}}
\end{pmatrix}$
\end{center}
The charged vector boson mass matrix can be written as

\begin{equation}
M_{V^\pm}^2=\left(
\begin{array}{cc}
 \frac{1}{4} v_L^2 g_L^2+\frac{1}{4} \left(k_1^2+k_2^2\right) g_L^2 & -\frac{1}{2} g_L g_R k_1 k_2 \\
 -\frac{1}{2} g_L g_R k_1 k_2 & \frac{1}{4} v_R^2 g_R^2+\frac{1}{4} \left(k_1^2+k_2^2\right) g_R^2 \\
\end{array}
\right)
\end{equation}
whereas the neutral vector boson mass matrix is

\begin{equation}
M_{V^0}^2 = \left(
\begin{array}{ccc}
 \frac{1}{8} g_L^2 \left(k_1^2+k_2^2+v_L^2\right) & -\frac{1}{4} g_L g_R \left(k_1^2+k_2^2\right) & -\frac{1}{4}g_{\text{BL}} g_L  v_L^2 \\
 -\frac{1}{4} g_L g_R \left(k_1^2+k_2^2\right) & \frac{1}{8} g_L^2 \left(k_1^2+k_2^2+v_R^2\right) & -\frac{1}{4}g_{\text{BL}} g_R  v_R^2 \\
 -\frac{1}{4}g_{\text{BL}} g_L  v_L^2 & -\frac{1}{4}g_{\text{BL}} g_R  v_R^2 & \frac{1}{8} g_{\text{BL}}^2  \left(v_L^2+v_R^2\right) \\
\end{array}
\right)
\end{equation}
As expected, the neutral gauge boson mass matrix has one vanishing eigenvalue, corresponding to massless photon. After diagonalisation of the mass matrices we can represent the gauge fields in terms of physical gauge boson states as

\begin{eqnarray}\nonumber
W_{L_\mu}^3 &=& \frac{e}{g_Y} Z_{L\mu} + \frac{e}{g_L} A_{\mu} + 0 \,\, Z_{R\mu}\\ \nonumber
W_{R_\mu}^3 &=& -\frac{e g_Y}{g_L g_R} Z_{L\mu} + \frac{e}{g_R} A_{\mu} + \frac{g_Y}{g_{BL}} \,\, Z_{R\mu}\\
B^{\prime}_{\mu} &=& -\frac{e g_Y}{g_L g_{BL}} Z_{L\mu} + \frac{e}{g_{BL}} A_{\mu} - \frac{g_Y}{g_R} \,\, Z_{R\mu}
\label{neutral:basis1}
\end{eqnarray} 
Also, we can express these couplings as, $\text{sin}\theta_W = \frac{e}{g_L}$ and $\text{cos}\theta_W = \frac{e}{g_Y}$ with $\theta_W$ being the Weinberg angle. In DLRM, $Z_{L\mu}$ and $Z_{R\mu}$ will also mix as the bi-doublet $\Phi$ transform non-trivially under both $SU(2)_L$ and $SU(2)_{R}$ gauge groups. The mixing can be represented as 

\begin{eqnarray}\nonumber 
Z_{L\mu} &= & \cos \delta \,\, Z_\mu - \sin \delta \,\, Z^\prime_\mu\\
Z_{R\mu} &= & \sin \delta \,\, Z_\mu + \cos \delta \,\, Z^\prime_\mu
\label{neutral:basis2}
\end{eqnarray} 
where the mixing angle can be written as
\begin{equation}
\tan 2 \delta = \frac{2 (M_{L,R}^{0})^2}{(M_L^0)^2-(M_R^0)^2}
\label{tan:2delta}
\end{equation}
with  
\begin{eqnarray}
(M_L^0)^2 &=& \frac{e^2(g_L^2+g_Y^2)^2(k_1^2+k_2^2+v_L^2)}{8g_L^2g_Y^2} \\
(M_{L,R}^{0})^2 &=& \frac{e(g_L^2+g_Y^2)(g_R^2(k_1^2+k_2^2)-g_{BL}^2v_L^2)}{4 g_{BL}g_L g_R} \\
(M_R^0)^2 &=& \frac{g_Y^2(2g_{BL}^2g_R^2v_R^2+g_R^4(k_1^2+k_2^2+v_R^2)+g_{BL}^4(v_L^2+v_R^2))}{8g_{BL}^2g_R^2}
\end{eqnarray}
The charged vector boson states are
\begin{eqnarray}\nonumber
W_{L\mu}^\pm &= & \cos \zeta \,\, W_{1\mu}^\pm - \sin \zeta \,\, W_{2\mu}^\pm\\ 
W_{R\mu}^\pm &= & \sin \zeta \,\, W_{1\mu}^\pm + \cos \zeta \,\, W_{2\mu}^\pm
\label{phy:basis:WLRpm}
\end{eqnarray}
with
\begin{equation}
\tan 2 \zeta = \frac{2 M_{LR}^2}{M_{L}^2 - M_{R}^2}
\end{equation}
where,
\begin{eqnarray}
M_{L}^2 &=&\frac{1}{4}g_L^2 (k_1^2+k_2^2+v_L^2)\\
M_{R}^2 & = & \frac{1}{4}g_R^2 (k_1^2+k_2^2+v_R^2) \\
M_{LR}^2 & = & -\frac{1}{2}g_Lg_Rk_1k_2
\end{eqnarray}
After diagonalisation of $\lbrace W_{L}^\pm , W_R^\pm \rbrace $ to $\lbrace W_{1}^\pm , W_2^\pm \rbrace $, we can have the corresponding mass-squared terms for the charged physical gauge bosons as,
\begin{eqnarray}
M_{1}^2 &=& \frac{1}{4}\left(g_L^2 (k_1^2+k_2^2+v_L^2)\text{cos}^2\zeta -2g_L g_R k_1 k_2 \text{sin}2\zeta + g_R^2 (k_1^2+k_2^2+v_R^2)\text{sin}^2\zeta \right) \\
M_{2}^2 &=& \frac{1}{4}\left(g_R^2 (k_1^2+k_2^2+v_R^2)\text{cos}^2\zeta +2g_L g_R k_1 k_2 \text{sin}2\zeta + g_L^2 (k_1^2+k_2^2+v_L^2)\text{sin}^2\zeta \right)
\end{eqnarray}
Note that we have taken $k_2 =0$ which is equivalent to vanishing tree level mixing angle $\zeta$. One can however generate radiative mixing between charged vector bosons, but that is typically very small $< 10^{-7}$ \cite{Borah:2017leo}. Note that although we write $W_1, W_2, Z, Z^{\prime}$ as physical massive gauge boson states here to show the details, in the main text we continue to use $W_L, W_R, Z_L, Z_R$ for better clarity.

\section{Fermion-gauge boson interactions in DLRM}
\label{appen2}
In this section we note down the fermion interactions with massive vector bosons. The kinetic term of leptons in DLRM is given by

\begin{eqnarray}
\mathcal{L}_{\ell}=i\overline{\ell_L} \slashed{D}^L \ell_L + i\overline{\ell_R} \slashed{D}^R \ell_R
\end{eqnarray}
where,

\begin{eqnarray}
D^L_{\mu} \ell_L &=& \partial_{\mu}\, \ell_{L} - i \frac{g_L}{2} (\sigma. W_{L\mu})\ell_L + i \,\, \frac{g_{BL}}{2}\, B_{\mu}\ell_L\\
D^R_{\mu} \ell_R &=&  \partial_{\mu}\, \ell_{R} - i \frac{g_R}{2} (\sigma. W_{R\mu})\ell_R + i \,\, \frac{g_{BL}}{2}\, B_{\mu}\ell_R
\end{eqnarray}
The same kinetic Lagrangian is, in fact, applicable to quarks too if we include gluons in the covariant derivative.
We show the interactions with neutral massive vector bosons in table \ref{tab:2B}, \ref{tab:2C} respectively.
\begin{table}[H]
\centering
\renewcommand{\arraystretch}{1.4}
\begin{tabular}{|p{2cm}||p{5cm}|p{10cm}|}
 \hline
     \multirow{2}{*}{Fermions} &
 \multicolumn{2}{|c|}{Z Bosons}  \\
 \cline{2-3}

 &  \hspace{1.5cm} {\rm $Z\, \text{in  SM}$} &\hspace{4cm} {$Z \, \text{in DLRM}$}\\
\hline
${\rm \overline{e_L}\gamma_\mu e_L}$ & $-i\frac{g_L}{2\cos\theta_W} \cos2\theta_W $ &$i\left(-\frac{g_L \text{cos}2\theta_W}{2\text{cos}\theta_W}\text{cos}\delta + \frac{g_{BL}g_L \text{tan}\theta_W \text{sin}\delta}{2g_R} \right)$ \\ 
\hline
${\rm \overline{e_R}\gamma_\mu e_R}$ & $i\frac{g_L}{\cos\theta_W}\sin^2\theta_W$  & $i \left( \frac{g_L \text{sin}^2 \theta_W \text{cos}\delta}{\text{cos}\theta_W} +\frac{1}{2} \left(\frac{g_{BL}}{g_R}-\frac{g_R}{g_{BL}}\right) g_L \text{tan}\theta_W \text{sin}\delta  \right)$\\ 
\hline
${\rm \overline{\nu_L}\gamma_\mu \nu_L}$ & $i\frac{g_L}{2\cos\theta_W}$  &$i \left(\frac{g_L (1-2\text{sin}^2\theta_W \text{cos}^2\theta_W)\text{cos}\delta}{2\text{cos}^2\theta_W \text{sin}\theta_W} + \frac{g_{BL}g_L \text{tan}\theta_W \text{sin}\delta}{2g_R}\right)$\\ 
\hline
${\rm \overline{u_L}\gamma_\mu u_L}$ & $-i \frac{g_L}{2\cos\theta_W}\left( \cos^2\theta_W -\frac{\sin^2\theta_W}{3}\right)$ &$i \left(-g_L \text{sin}\theta_W \left(\frac{\text{cot}\theta_W}{2}-\frac{\text{tan}\theta_W}{6}\right)\text{cos}\delta + \frac{g_{BL}g_L \text{tan}\theta_W \text{sin}\delta}{6g_R} \right)$ \\ 
\hline
${\rm \overline{u_R}\gamma_\mu u_R}$ & $-i \frac{2}{3}g_L \frac{\sin^2\theta_W}{\cos\theta_W}$ &$i \left( \frac{1}{6}g_L \text{tan}\theta_W \left(-4\text{sin}\theta_W \text{cos}\delta + \left(-\frac{g_{BL}}{g_R}+\frac{3g_R}{g_{BL}}\right) \text{sin}\delta \right)  \right)$ \\ 
\hline
${\rm \overline{d_L}\gamma_\mu d_L}$ &$-i\frac{g_L}{2\cos\theta_W}$  &$i \left(-\left(\frac{g_L \text{cos}\theta_W}{2}+\frac{g_L \text{sin}^2 \theta_W}{6 \text{cos}\theta_W}\right)\text{cos}\delta - \frac{g_{BL}g_L \text{tan}\theta_W \text{sin}\delta}{6g_R}\right)$ \\ 
\hline
${\rm \overline{d_R}\gamma_\mu d_R}$ & $i\frac{1}{3}g_L \frac{\sin^2\theta_W}{\cos\theta_W}$  &$i \left( \frac{1}{6}g_L \text{tan}\theta_W \left(2\text{sin}\theta_W \text{cos}\delta +\newline \left(-\frac{g_{BL}}{g_R}-\frac{3g_R}{g_{BL}}\right) \text{sin}\delta \right)  \right)$\\ 
\hline
${\rm \overline{\nu_R} \gamma_\mu \nu_R}$ & 0 & $i \left(\frac{(g_{BL}^2+g_R^2)g_L \text{tan}\theta_W \text{sin}\delta}{2g_{BL}g_R}\right)$\\
\hline
\end{tabular}
\caption{Fermion interactions with Z boson.}
\label{tab:2B}
\end{table}

\begin{table}
\centering
\renewcommand{\arraystretch}{1.4}
\begin{tabular}{|p{2cm}||p{10cm}|}
 \hline
 Fermions &\hspace{3.5cm} {\rm $Z^\prime\, \text{in  LRSM}$} \\
\hline
${\rm \overline{e_L}\gamma_\mu e_L}$ & $i\left(\frac{g_{BL}g_L \text{tan}\theta_W \text{cos}\delta}{2g_R} + \frac{g_L \text{cos}2\theta_W \text{sin}\delta}{\text{cos}\theta_W}\right)$ \\
\hline
${\rm \overline{e_R}\gamma_\mu e_R}$ &  $i\left(\frac{1}{2}g_L \text{tan}\theta_W \left(\left(\frac{g_{BL}}{g_R}-\frac{g_R}{g_{BL}}\right)\text{cos}\delta -2\text{sin}\theta_W \text{sin}\delta \right)\right)$\\
\hline
${\rm \overline{\nu_L}\gamma_\mu \nu_L}$ &$i \left(\frac{g_{BL}g_L \text{tan}\theta_W \text{cos}\delta}{2g_R}-\frac{g_L (1-2\text{sin}^2\theta_W \text{cos}^2\theta_W)\text{sin}\delta}{2\text{cos}^2\theta_W \text{sin}\theta_W}\right)$ \\
\hline
${\rm \overline{\nu_R}\gamma_\mu \nu_R}$ & $i \left(\frac{(g_{BL}^2+g_R^2)g_L \text{tan}\theta_W \text{cos}\delta}{2g_{BL} g_R}\right)$\\
\hline
${\rm \overline{u_L}\gamma_\mu u_L}$ & $i \left(-\frac{g_{BL}g_L \text{tan}\theta_W \text{cos}\delta}{6g_R}+g_L \text{sin}\theta_W \left(-\frac{\text{cot} \theta_W}{2}+\frac{\text{tan}\theta_W}{6}\right)\text{sin}\delta \right)$\\
\hline
${\rm \overline{u_R}\gamma_\mu u_R}$ & $i \left(\frac{1}{6}g_L \text{tan}\theta_W \left(-\frac{g_{BL}}{g_R}+\frac{3g_R}{g_{BL}}\right)\text{cos}\delta + 4\text{sin}\theta_W \text{sin}\delta \right)$\\
\hline
${\rm \overline{d_L}\gamma_\mu d_L}$ & $i \left(-\frac{g_{BL}g_L \text{tan}\theta_W \text{cos}\delta}{6g_R}+g_L \text{sin}\theta_W \left(\frac{\text{cot}\theta_W}{2}+\frac{\text{tan}\theta_W}{6}\right)\text{sin}\delta \right)$\\
\hline
${\rm \overline{d_R}\gamma_\mu d_R}$ & $i \left(-g_L \text{tan}\theta_W \left(\frac{(g_{BL}^2+3g_R^2)\text{cos}\delta}{6g_{BL}g_R}+\frac{\text{sin}\theta_W \text{sin}\delta}{3}\right)\right)$\\
\hline
\end{tabular}
\caption{Fermion interaction with $Z^{\prime}$ boson.}
\label{tab:2C}
\end{table}

The interaction of neutrino $\nu$, charged leptons $\ell$ with $W_2$ (or $W_R$) is similar to the ones with $W_L$ except that $g_L$ is replaced by $g_R$:
$$ -\frac{i g_R}{\sqrt{2}} \gamma_\mu P_R$$
where we have ignored the details of right handed lepton mixing matrix, taking it to be a unit matrix.
\section{Annihilation cross-sections of right handed neutrinos}
\label{appen3}
The annihilation cross sections of $\nu_R$ mediated by right sector gauge bosons are 
\begin{align}
\sigma_{\nu_R \overline{\nu_R}\rightarrow q \overline{q}} = \frac{ \left(a^2+b^2\right) \sqrt{1-\frac{4 m_q^2}{s}} \left(c^2 \left(2 m_q^2+s\right)+d^2 \left(s-4 m_q^2\right)\right)}{192 \pi  M_{Z_R}^4} 
\end{align}
\begin{align}
\sigma_{\nu_R \overline{\nu_R}\rightarrow \ell_R \overline{\ell_R}}  &= \frac{\sqrt{1-\frac{4 m_\ell^2}{s}}}{192 \pi  M_{W_R}^4 M_{Z_R}^4} \bigg(2 m_\ell^2 \big(M_{W_R}^4 \left(a^2+b^2\right) \left(c^2-2 d^2\right)- M_{W_R}^2 M_{Z_R}^2 (a+b)(c-2 d)- \nonumber \\
& M_{Z_R}^4 (a+b)\big)+s \big(M_{W_R}^4 \left(a^2+b^2\right) \left(c^2+d^2\right) \nonumber \\
& -M_{W_R}^2 M_{Z_R}^2 (a+b) (c+d)+2 M_{Z_R}^4 (a+b)\big)\bigg)
\end{align}
where
\begin{equation}
a=b=\frac{g_L\,\tan \theta_W}{2}\frac{g_{BL}^2+g_R^2}{2 g_{BL}\, g_R}\\
\end{equation}
\begin{equation}
c=
\begin{cases}
\frac{g_L\,\tan \theta_W}{2} \frac{2g_{BL}^2-g_R^2}{2 g_{BL} g_R} & \text{(for charged leptons)} \\
\frac{g_L\,\tan \theta_W}{2} \frac{3g_{R}^2-2g_{BL}^2}{6 g_{BL} g_R} & \text{(for up type quark)} \\
-\frac{g_L\,\tan \theta_W}{2} \frac{2g_{BL}^2+g_R^2}{2 g_{BL} g_R} & \text{(for down type quark)} 
\end{cases}
\end{equation}
\begin{equation}
d=
\begin{cases}
-\frac{g_L\,\tan \theta_W}{4 g_{BL}} g_R & \text{(for charged leptons)}\\
\frac{g_L\,\tan \theta_W}{4 g_{BL}}g_R  & \text{(for up type quark)}\\ 
-\frac{g_L\,\tan \theta_W}{4 g_{BL}} g_R & \text{(for down type quarks)}
\end{cases}
\end{equation}
We consider the mixing between left and right sector gauge bosons to be negligible and hence do not take it into account in our analysis.


\providecommand{\href}[2]{#2}\begingroup\raggedright\endgroup

\end{document}